\documentclass[letterpaper,11pt]{article}
\usepackage{amsthm}  
\usepackage{amsmath, amssymb}
\usepackage{color}
\usepackage{fullpage}
\usepackage[numbers]{natbib}
\usepackage[noend]{algorithmic}
\usepackage{thmtools, thm-restate}
\usepackage{tikz}
\usepackage{enumerate}
\usepackage{thm-restate}
\usepackage{graphicx}
\usepackage{amsopn}
\usepackage{caption}
\usepackage{hyperref}
\usepackage{subcaption}
\usepackage{zerosum,csquotes}
\usepackage{enumitem,linegoal}
\usepackage[linesnumbered,ruled,vlined]{algorithm2e}
\usepackage{comment}
\usepackage{float}

\usepackage{ctable}					

\usepackage{mathtools}
\usepackage[flushleft]{threeparttable}
\usepackage[normalem]{ulem}

\usepackage{footnote}
\makesavenoteenv{tabular}
\makesavenoteenv{table}

\usepackage[margin=1in]{geometry}

 \newtheorem{theorem}{Theorem}[section]
 \newtheorem{lemma}[theorem]{Lemma}

\theoremstyle{definition}

\makeatletter
\newif\ifqed
\def\GrabProofArgument[#1]{ #1: \egroup\ignorespaces}
\def\proof{\noindent\textbf\bgroup Proof%
	\@ifnextchar[{\GrabProofArgument}{. \egroup\ignorespaces}\global\qedtrue}

\def\qedhere{\ifmmode\tag*{\qedsign}\else\hspace*{\fill}\qedsign\medskip\fi\global\qedfalse}
\def\qedsign{$\Box$}
\makeatother

\newcommand{\match}{\mathcal{M}}
\newcommand{\ms}{\mu}
\DeclareMathOperator{\alg}{ALG}
\DeclareMathOperator{\pl}{polylog}
\DeclareMathOperator{\gr}{GREEDY}
\DeclareMathOperator{\bm}{BIPARTITE-MATCHING}
\DeclareMathOperator{\barg}{BIPARTITE-AUG-PATH}
\DeclareMathOperator{\farg}{BIPARTITE-5AUG-PATH}

\DeclareMathOperator{\gm}{MATCHING-VIA-BIPARTITE}
\newcommand{\E}{\mathbb{E}}

\newcommand{\aliz}[1]
{\par {\color{blue} Alireza: #1 \par}}

\definecolor{mygreen}{RGB}{20,140,80}
\definecolor{mylightgray}{RGB}{230,230,230}

\definecolor{mygreen}{RGB}{20,140,80}
\definecolor{mydarkgray}{gray}{0.15} 
\definecolor{oceanblue}{HTML}{2c55c2}
\hypersetup{
     colorlinks=true,
     citecolor= mygreen,
     linkcolor= black
}

\newcommand{\citeboth}[1]{\hypersetup{citecolor=mydarkgray}\citeauthor{#1}\hypersetup{citecolor=mygreen} \cite{#1}}

\SetCommentSty{mycommfont}

\newcommand*\samethanks[1][\value{footnote}]{\footnotemark[#1]}

\newcounter{proccnt}

\newcommand{\konote}[1]{}

\title{Approximate Maximum Matching in Random Streams}
\author{
	Alireza Farhadi\thanks{University of Maryland. Email: \texttt{\{farhadi,hajiagha\}@cs.umd.edu}.}
	\and
	MohammadTaghi Hajiaghayi\samethanks[1]
	\and
	Tung Mai\thanks{Adobe Research. Email: \texttt{\{tumai,anuprao,rrossi\}@adobe.com}.}
	\and
	Anup Rao\samethanks[2]
	\qquad\quad
	Ryan A. Rossi\samethanks[2]
}

\begin{document}
	\newcommand{\ignore}[1]{}
\renewcommand{\theenumi}{(\roman{enumi})}
\renewcommand{\labelenumi}{\theenumi.}
\sloppy

%
%

\date{}

\maketitle


\begin{abstract}
In this paper, we study the problem of finding a maximum matching in the semi-streaming model when edges arrive in a random order. In the semi-streaming model, an algorithm receives a stream of edges and it is allowed to have a memory of $\tilde{O}(n)$\footnote{We use $\tilde{O}(.)$ notion to hide logarithmic factors.} where $n$ is the number of vertices in the graph. A recent inspiring work by  \citeboth{DBLP:conf/soda/AssadiBBMS19} shows that there exists a streaming algorithm with the approximation ratio of $\frac{2}{3}$ that uses $\tilde{O}(n^{1.5})$ memory. However, the memory of their algorithm is much larger than the memory constraint of the semi-streaming algorithms. In this work, we further investigate this problem in the semi-streaming model, and 
we present simple algorithms for approximating maximum matching in the semi-streaming model.
Our main results are as follows.
\begin{itemize}
\item We show that there exists a single-pass deterministic semi-streaming algorithm that finds a $\frac{3}{5} (= 0.6)$ approximation of the maximum matching in bipartite graphs using $\tilde{O}(n)$ memory. This result significantly outperforms the state-of-the-art result of \citeboth{DBLP:conf/mfcs/Konrad18} that finds a $0.539$ approximation of the maximum matching using $\tilde{O}(n)$ memory.
\item By giving a black-box reduction from finding a matching in general graphs to finding a matching in bipartite graphs, we show there exists a single-pass deterministic semi-streaming algorithm that finds a $\frac{6}{11} (\approx 0.545)$ approximation of the maximum matching in general graphs, improving upon the state-of-art result $0.506$ approximation by \citeboth{DBLP:journals/corr/abs-1811-02760}. 
\end{itemize}

\end{abstract}
\section{Introduction}
As noted by Lov{\'a}sz and Plummer in their book \cite{plummer1986matching} ``Matching Theory is a central part of graph theory, not only because of its applications, but also because it is the source of important ideas developed during the rapid growth of combinatorics during the last several decades". Classical algorithms for finding a matching assume that we have enough memory to store all of the vertices and edges of the graph in the memory. However, in the era of big data, we often face with massive datasets that are too large to fit into the computer's memory. \textit{Semi-streaming} is one of the models introduced to deal with massive inputs \cite{DBLP:conf/soda/FeigenbaumKMSZ05}. In this model, the algorithm has access to a sequence of edges and it is allowed to have a memory of $\tilde{O}(n)$ where $n$ is the number of vertices in the graph. Finding a large matching as a fundamental problem in graph theory has been extensively studied in the semi-streaming model.

When edges of the graph arrive in an arbitrary order, it is known that a simple algorithm that adds edges to the matching in a greedy manner, can find a maximal matching, and in fact, $1/2$ approximation of the maximum matching using $\tilde{O}(n)$ memory. However, whether or not we can get better than $1/2$ approximation with $\tilde{O}(n^{2-\epsilon})$ memory for a constant $\epsilon >0$ has remained an important open question since the introduction of the semi-streaming model \cite{DBLP:conf/soda/FeigenbaumKMSZ05}. It is shown in a series of works \cite{DBLP:conf/soda/GoelKK12, DBLP:conf/soda/Kapralov13} that there is no semi-streaming algorithm for the maximum matching problem with a approximation ratio better than $1-1/e$. Specifically, a work by \citeboth{DBLP:conf/soda/Kapralov13} shows that any streaming algorithm with the approximation ratio of better than $1-1/e$ requires $n^{1+\Omega(1/\log \log n)}$ memory. However, this still leaves room between $1/2$ and $1-1/e$. In order to make progress on this long-standing open problem, an influential work by \citeboth{DBLP:conf/approx/KonradMM12} studied the problem when edges arrive in a uniformly random order. Despite the fact that the approximation ratio of the greedy algorithm could be $1/2 +o(1)$ in this setting, they surprisingly showed there exists an algorithm that achieves the approximation ratio of $1/2 + 0.003$ using $\tilde{O}(n)$ memory. A recent inspiring work of \citeboth{DBLP:conf/soda/AssadiBBMS19} uses \textit{edge degree constrained subgraphs} (EDCS) introduced by \citeboth{DBLP:conf/icalp/BernsteinS15} to design streaming algorithms and they show that there exists a streaming algorithm for the maximum matching problem with the approximation ratio of $2/3$ that uses $\tilde{O}(n^{1.5})$ memory. However, the memory of their algorithm is not within memory bounds of semi-streaming algorithms.

\begin{table}
\centering
\setlength{\tabcolsep}{12pt}
\begin{tabular}{l llc} 
\toprule
& \multicolumn{2}{c}{\textsc{Approximation Factor}} & \\
& \textbf{Bipartite graphs} & \textbf{General graphs} & \textsc{Space} \\
\midrule

\citeboth{DBLP:conf/soda/KapralovKS14} & $\frac{1}{\pl n}$ & $\frac{1}{\pl n}$ & $O(\pl n)$ \\
\midrule

\citeboth{DBLP:conf/approx/KonradMM12} & $0.5005$ & $0.5003$ & $\tilde{O}(n)$ \\

\citeboth{DBLP:journals/corr/abs-1811-02760} & $0.512$ & $0.506$ & $\tilde{O}(n)$ \\
 
\citeboth{DBLP:conf/mfcs/Konrad18} & $0.539$ & - & $\tilde{O}(n)$ \\

\textbf{This paper} & $\mathbf{0.6}$ & $\mathbf{0.545}$ & $\tilde{O}(n)$ \\

\bottomrule
\end{tabular}
\caption{Single-pass semi-streaming algorithms known for the maximum matching when edges arrive in a random order.}
\label{table:relatedwork}
\vspace{-0.1cm}
\end{table}

In this work, we further study the problem of finding a maximum matching in random arrival of edges. In this model, \citeboth{DBLP:conf/approx/KonradMM12} showed that the $1/2$ barrier can be broken, and they designed a semi-streaming algorithm with the approximation ratio of $0.5003$ in general graphs and $0.5005$ in bipartite graphs. These approximation factors are later improved to $0.539$ in bipartite graphs and $0.506$ in general graphs \cite{DBLP:journals/corr/abs-1811-02760,DBLP:conf/mfcs/Konrad18}. Also, a surprising result by \citeboth{DBLP:conf/soda/KapralovKS14} showed that when edges arrive in a random order, there exists a streaming algorithm that finds $(1/\pl n)$ approximation of the size of a maximum matching using only poly-logarithmic space. (Table \ref{table:relatedwork}) gives a brief survey of known results in a random stream of edges as well as a comparison to our results.

We improve the state-of-the-art results in both bipartite and general graphs when edges arrive in a random order. We show that there exists a single-pass semi-streaming algorithm that finds a $0.6$ approximation of the maximum matching in bipartite graphs.  

\begin{restatable}{theorem}{mainone}
\label{thm:m1}
There exists a deterministic single-pass semi-streaming algorithm that w.h.p.\footnote{With high probability} finds a $\big(\frac{3}{5} -o(1)\big)$ approximate of the maximum matching in bipartite graphs when edges arrive in a uniformly random order. 
\end{restatable}
This result significantly improves the randomized algorithm of \cite{DBLP:conf/mfcs/Konrad18} that finds a $0.539$ approximation of the maximum matching in bipartite graphs. We also give a black-box reduction from finding a matching in general graphs in the semi-streaming model to the problem of finding a matching in bipartite graphs. Our reduction result is as follows.

\begin{restatable}{theorem}{maintwo}
\label{thm:m2}
Given any single-pass semi-streaming algorithm with the approximation ratio of $p$ for finding a maximum matching in bipartite graphs in random arrival of edges, there exists a semi-streaming algorithm that finds w.h.p. a $\big(\frac{2p}{2p+1}-o(1)\big)$ approximation of the maximum matching when edges arrive in a random order.
\end{restatable}

Using this reduction and Theorem \ref{thm:m1}, we can immediately get the following result.
\begin{restatable}{theorem}{mainthree}
\label{thm:m3}
There exists a deterministic single-pass semi-streaming algorithm that finds w.h.p. a $\big(\frac{6}{11} -o(1)\big)(\approx 0.545)$ approximation of the maximum matching in general graphs when edges arrive in a random order. 
\end{restatable}

\subsection{Related Work}
In case of weighted graphs, the problem of finding a maximum weighted matching first considered by \citeboth{DBLP:journals/tcs/FeigenbaumKMSZ05}. They showed that a modification of the greedy algorithm can find a $1/6$ approximation of the maximum weighted matching in the semi-streaming model. 
This bound later improved in a series of works \cite{DBLP:conf/approx/McGregor05,DBLP:journals/algorithmica/Zelke12,DBLP:journals/siamdm/EpsteinLMS11,DBLP:conf/approx/CrouchS14,DBLP:conf/soda/PazS17,DBLP:conf/soda/GhaffariW19} to $(1/2 - \epsilon)$ approximation. A recent work by \citeboth{DBLP:journals/corr/abs-1811-02760} shows that when edges arrive in a random order, there exists a semi-streaming algorithm that finds a $\big(1/2+ \Omega(1)\big)$ approximation of the maximum weighted matching, beating the $1/2$ barrier for the weighted case.

Another line of research considers the problem of finding an approximate matching in the semi-streaming model using a very small number of passes \cite{DBLP:conf/icdm/EsfandiariHM16, DBLP:conf/approx/KaleT17}.

\subsection{Overview of Our Algorithm}
Our algorithm starts by using a very small fraction of the edges and gives these edges to the greedy algorithm to find a matching $\match_0$. Let $R$ be the set of edges in the graph that do not share a vertex with the edges in $\match_0$. Since the order of edges is uniformly random, it can be shown that (formalized as in Lemma \ref{lem:smallgreedy}) the number of edges in $R$ should be small and we can fit all of them in the semi-streaming memory of our algorithm. Thus, we can find the maximum matching of $R$ using any offline algorithm. For now, just suppose that we want to break the $0.5$-approximation barrier. Let $\ms$ be the size of maximum matching in the graph. There are at most $2|\match_0|$ matching edges that share a vertex with the edges in $\match_0$. Therefore, the size of the maximum matching in $R$ is at least $\ms -2|\match_0|$. Combining this matching with the matching $\match_0$ gives us a matching of size at least $\ms -|\match_0|$. Therefore, we can get a matching of size $\max \{|\match_0|, \ms -|\match_0| \}$ using this simple algorithm. The only case that we do not break the $0.5$-approximation barrier is when we roughly have $|\match_0| = \ms/2$. Also, the size of a maximum matching in $R$ should be almost zero. This implies that $\match_0$ is almost a maximal matching.

Therefore, in the worst case scenario of the mentioned algorithm, the algorithm can find a maximal matching $\match_0$ by looking at a very small fraction of the edges. Since $|\match_0| = \ms/2$, there are $|\match_0|$ edge disjoint $3$-augmenting paths for the edges in $\match_0$. All of the previous works in the semi-streaming model use the remaining edges in the streaming to find augmenting paths for $\match_0$ and increase the size of the matching. Our work is different from them from  several perspectives.

\begin{itemize}
\item All of the previous works, only use the fact that order of the edges is random to argue that the size of $\match_0$ is very close to the half of the size of a maximum matching. For finding augmenting paths, they use algorithms that also work when order of edges is adversarial. However, finding augmenting paths is much trickier than finding a matching in graph, and in fact, finding a matching in a graph can easily be reduced to finding augmenting paths. It seems the assumption that order of edges is uniformly at random makes the problem of finding a matching easier. Therefore, this assumption should also helps us to find a better set of augmenting paths. Our algorithm is a recursive algorithm that extensively uses the fact that the order of edges is random, and it finds much better set of augmenting paths.

\item Unlike the previous works on this problem, our algorithm finds augmenting paths with a length larger than $3$. We first propose an algorithm for finding $3$-augmenting paths, and we show that the approximation ratio of this algorithm is roughly $4/7$ in bipartite graphs. We then modify this algorithm to find a small set of $5$-augmenting paths, and we show that using this modification, the approximation ratio of our algorithm improves to $3/5$.

\item The state-of-the-art algorithm for finding semi-streaming matching is randomized \cite{DBLP:conf/mfcs/Konrad18}, however our algorithm is deterministic and its expected performance only depends on the random order of the edges, not the random choices of the algorithm.
\end{itemize}
\section{Preliminaries}

Let $G=(V,E)$ be a graph. A subset of edges $\match \subseteq E$ is a matching if no two different edges in $\match$ share a same endpoint. For any set of edges $E$, we use $\ms(E)$ to denote the size of a maximum matching in $E$. We may also abuse the notation through  the paper and use $\ms(E)$ to refer to the set of edges in a maximum matching. We use $\ms(G)$ to denote the matching of $G$. We use $V(G)$ and $E(G)$ to denote the set of vertices and edges of a graph $G$, respectively. For a subgraph $S \subseteq G$, we use $\overline{V(S)}$ to denote the vertices in $G$ that are not in $V(S)$, i.e., $\overline{V(S)} = V(G) \setminus V(S)$. In case that $G=(A,B,E)$ is a bipartite graph, we use $\overline{A(S)}$ (respectively,  $\overline{B(S)}$) to denote the vertices in $A$ (resp., $B$) that are not in $V(S)$.
When $G=(A,B,E)$ is a bipartite graph, for any edge $e=(a,b) \in E$, we always assume that $a \in A$ and $b \in B$.

For a graph $G$, let $m=|E(G)|$ be the number of edges in $G$. A graph stream is a sequence of distinct edges $S=\langle e_1, \ldots, e_{m} \rangle$ where $(e_1,\cdots, e_m)$ is a permutation of the edges. Throughout this paper, we assume that this permutation is chosen uniformly at random from all permutations. We use $S_{[a,b]}$ to denote the stream $\langle e_a, \cdots, e_b \rangle$. Also, we use $S_{(a,b]}$ to denote the stream $\langle e_{a+1}, \cdots, e_b \rangle$.  A simple greedy algorithm (Algorithm \ref{alg:greedy-unweighted}) returns a $1/2$-approximation of the maximum matching when order of edges is arbitrary. However, it is easy to see that the approximation ratio of the greedy algorithm when edges arrive in a random order could be $1/2+o(1)$. For example, consider a graph $G=(A,B,E)$ where $|A|=|B|=n$. Let $A=\{a_1, \cdots, a_n\}$ and $B=\{b_1, \cdots, b_n \}$. Then connect $a_i$ to $b_j$ if $i=j$ or $i\le n/2$ and  $j> n/2$. Although $G$ has a perfect matching, the subgraph between the first $n/2$ vertices of $A$ and the last $n/2$ vertices of $B$ has $\Omega(n^2)$ edges, and the algorithm is very likely to pick these edges in the matching.

\begin{algorithm} [h]
 \KwData{An arbitrary stream $S=\langle e_1, \ldots, e_m \rangle$ of edges of a graph $G=(V,E)$.}
 \begin{algorithmic} [1]
 \STATE  Initially set $\match = \O$. 
 \FOR {each $e \in S$}
    \IF {$\match \cup \{e\}$ is a matching}
        \STATE $\match \leftarrow \match \cup \{e\}$.
    \ENDIF
 \ENDFOR
 \RETURN $\match$.
 \end{algorithmic}
\caption{Algorithm $\gr$.}
\label{alg:greedy-unweighted}
\end{algorithm}

A result by \citeboth{DBLP:conf/soda/ChitnisCEHMMV16} shows that the maximum matching can be found in the streaming model with a memory of size $\tilde{O}(k^2)$ where $k$ is the cardinality of the maximum matching. Therefore, if $\ms(G) \le \sqrt{n}$, using their algorithm we can find the exact solution using a memory of size $\tilde{O}(n)$. In fact, we can always run their algorithm in parallel, and set a limitation for the memory that it uses, and if $\ms(G)$ is small, this algorithm finds the exact solution. Therefore, throughout this paper we can assume that $\ms(G) \ge \sqrt{n}$.

In our analyses, we often use the Chernoff bound to prove concentration bounds (We refer the reader to \cite{wajc2017negative} for the an introduction to negative association and Chernoff bound).

\begin{theorem} [Chernoff bound]
Given $n$ negatively associated random variables $X_1, \cdots, X_n$ taking values in $\{0,1\}$, let $X$ denote their sum. Then, for $0 \le \delta \le 1$ we have
\begin{align*}
\Pr\big[ X \le (1-\delta) \E[X] \big] \le \exp\bigg(-\frac{\delta^2 \E[X]}{2}\bigg) \,.
\end{align*}
And,
\begin{align*}
\Pr\big[ X \ge (1+\delta) \E[X] \big] \le \exp\bigg(-\frac{\delta^2 \E[X]}{3}\bigg) \,.
\end{align*}

\end{theorem}

\section{Semi-Streaming Matching in Bipartite Graphs}

\begin{algorithm} [h]
 \KwData{A random order stream $S=\langle e_1, \ldots, e_m \rangle$ of edges of a bipartite graph $G=(A,B,E)$.}
 \begin{algorithmic} [1]
 \STATE Run $\gr$ on the edges $S_{[1,m/\log n]}$ to find a matching $\match_0$.
 \STATE Run the following subroutines in parallel.
 \STATE \hspace{\algorithmicindent} \textbf{Subroutine 1:}
 \STATE \hspace{\algorithmicindent}\hspace{\algorithmicindent} Run $\barg$ with the input of matching $\match_0$ and the stream of edges $e \in S_{(m/\log n,m]}$ such that exactly one of the endpoints of $e$ is in $V(\match_0)$. Let $T$ be the set of edges returned by $\barg$.
 \STATE \hspace{\algorithmicindent} \textbf{Subroutine 2:}

 \STATE \hspace{\algorithmicindent}\hspace{\algorithmicindent} Store all edges  $e \in S_{(m/\log n, m]}$ such that $e \cap V(\match_0) = \O$. Let $R$ be the set of these edges.
 \RETURN $\ms(\match_0 \cup T \cup R)$.
 \end{algorithmic}
\caption{Algorithm $\bm$ for finding an approximate matching in bipartite graphs.}
\label{alg:mainbipartite}

\end{algorithm}

In this section, we present our algorithm for finding an approximate matching in a random stream of edges. In our algorithm (formalized as in Algorithm \ref{alg:mainbipartite}), we first pick a small fraction of edges and find an approximate matching of these edges using $\gr$ algorithm. Specifically, we pick the first $1/\log n$ fraction of the edges in the stream and we use $\gr$ algorithm to find a matching $\match_0$. 
Let $R$ be the set of edges in the remaining stream that do not interfere with any of edges in $M_0$. In the following lemma which is first proved in \cite{DBLP:journals/corr/abs-1811-02760}, we show that w.h.p. the number of edges in $R$ is $\tilde{O}(n)$. Therefore, we can store all of them in the memory.

\begin{lemma}
\label{lem:smallremaining}
Let $G=(V,E)$ be a graph with $n$ vertices and $m$ edges, $H$ be a subgraph in $G$, and $S=\langle e_1,\cdots, e_m \rangle$ be a random stream of edges of $G$. Let $\match_0$ be the output of $\gr$ on edges of $S_{[1,\gamma \cdot m]}$ that are in $H$. Assume that $ \frac{1}{\log^{100} n} \le \gamma \le 1$, then w.h.p.,
$$
\big| E\big(H \setminus V(\match_0)\big) \big| = \tilde{O}\bigg(\dfrac{n}{\gamma}\bigg) \,.
$$
\end{lemma}

\begin{proof}
Let $m'=|E(H)|$ be the number of edges in $H$.
If $m' \le n$, then the lemma clearly holds. Otherwise assume that $m' > n$. We know that the expected number of edges of $E(H)$ that are in $S_{[1,\gamma \cdot m]}$ is $\gamma \cdot m'$. By Chernoff bound, the probability that less than $\gamma \cdot m'/2$ of these edges are in $S_{[1,\gamma \cdot m]}$, is at most
\begin{align*}
\exp\bigg(-\frac{\gamma \cdot m'}{8}\bigg) \le \exp\bigg(-\frac{n}{100 \log^{100} n}\bigg) 
\end{align*}
Which is less than $n^{-15}$ for a large enough $n$. Therefore, with a probability of at least $1 - n^{-15}$, the greedy algorithm sees $\gamma \cdot m'/2$ edges of $H$.
  
For a vertex $v$, Let $p_i$ be the probability of the event that $v$ is not matched using the first $i$ edges in $H$ and $v$ has at least $30 \log n / \gamma$ neighbors in $H$ among unmatched vertices. We show that $p_{\gamma \cdot m'/2} \le n^{-15}$. Therefore, by the union bound, the probability that at least one unmatched vertex has $30 \log n / \gamma$ neighbors in $H \setminus V(\match_0)$ is $n^{-14}$, and we can assume w.h.p. no such vertex exists, and it proves the lemma.

Consider the event that $v$ is not matched by the first $i-1$ edges and it has $d_i \ge 30 \log n / \gamma$ unmatched neighbors. In each step of $\gr$ algorithm we pick a random edge, and we add it to our matching if it do not interfere with the matching edges. The probability that $v$ remains unmatched in the step $i$ is at most $1- d_i / (m'-i+1)$. Since $d_i \ge 30 \log n / \gamma$, we have
$$
p_i \le p_{i-1} \bigg(1-\dfrac{30 \log n} {\gamma(m'-i+1)}\bigg) \le p_{i-1} \bigg(1-\dfrac{30 \log n} { \gamma m'}\bigg)   \,.
$$
Therefore,
$$
p_{\gamma \cdot m'/2} \le \prod_{i=1}^{\gamma \cdot m'/2} \bigg(1-\dfrac{30 \log n} {\gamma \cdot m'}\bigg) = \bigg(1-\dfrac{30 \log n} { \gamma \cdot m'}\bigg)^{\gamma \cdot m'/2} \le e^{-15 \log n} = n^{-15} \,.
$$
\end{proof}

According to the lemma above, the number of edges that do not interfere with any edge of $\match_0$ is small and w.h.p. we can store all of them in a memory of $\tilde{O}(n)$. Let $R$ be the set of these residual edges. Since edges in $R$ are disjoint from edges in $M_0$, we can compute the maximum matching in $R$ using any offline algorithm, and report $M_0 \cup \ms(R)$ as a matching. In the following lemma we show when size of $M_0$ is small, this approach gives us a good approximation of the maximum matching.

\begin{lemma}
\label{lem:smallgreedy}
Given a graph $G=(V,E)$, and a matching $\match$, let $R=E(G \setminus V(\match))$ be the set of edges that do not interfere edges in $\match$. If $|\match|= \alpha \ms(G)$, then $\ms(R) \ge (1-2 \alpha) \ms(G)$ and $|\match \cup \ms(R)| \ge (1-\alpha) \ms(G)$. 
\end{lemma}
\begin{proof}
Let $\match^*$ be an arbitrary maximum matching in $G$. Let $\match^*_1$ be the set of the edges in $\match^*$ that do not interfere with the edges in $\match$, i.e., $\match^*_1 = \match^* \cap R$, and let $\match^*_2 = \match^* \setminus \match^*_1$ be the set of edges in $\match^*$ that are incident to $V(\match)$. Since edges in $\match^*_2$ are a matching and they each are incident to a vertex in $V(\match)$, we have
\begin{align}
\label{lem:smallgreedy:eq1}
|\match^*_2| \le |V(\match)| \,.
\end{align}
Therefore,
\begin{align*}
|V(M)| \le 2 |\match| \,.
\end{align*}
This along with (\ref{lem:smallgreedy:eq1}) implies that
\begin{align*}
|\match^*_2| \le 2 |\match| = 2 \alpha \ms(G) \,.
\end{align*}
Therefore,
\begin{align*}
|\match^*_1| = |\match^*| - |\match^*_2| = \ms(G) -|\match^*_2| \ge \ms(G) (1-2\alpha) \,. 
\end{align*}
Note that $\ms(R)$ is the maximum matching in $G \setminus V(M)$, and all of the edges in $\match^*_1$ are in $G \setminus V(M)$. Therefore,
\begin{align}
\label{lem:smallgreedy:eq2}
\ms(R) \ge |\match^*_1| \ge \ms(G) (1-2\alpha) \,.
\end{align}
It implies,
\begin{align*}
|\match \cup \ms(R)| &= |\match| + |\ms(R)| \\ &
\ge |\match| + \ms(G)(1-2\alpha) & \text{By (\ref{lem:smallgreedy:eq2}).} \\
&= \alpha \ms(G) + \ms(G)(1-2\alpha) & \text{Since $|\match|=\alpha \ms(G)$.}
\\
&= \ms(G) (1-\alpha) \,,
\end{align*}
which proves the lemma.
\end{proof}

The mentioned lemma shows that if in our algorithm the size of the matching $\match_0$ is very small, then we can find a good approximate of the maximum matching by finding $\ms(R)$. Also, if size of the matching $\match_0$ is very large, we already have a large matching. Therefore, the hard case is when size of the $\match_0$ is neither too small nor too large.  In this case the algorithm calls $\barg$. Suppose that $E'$ is the set of edges in the remaining stream. The algorithm finds a set $3$-augmenting paths and returns a set of edges $T$ that are candidate edges for  $3$-augmenting paths. The algorithm $\barg$ guarantees to return a set of edges $T$ such that $\ms(T \cup R) \ge (\frac{4}{7} -o(1)) \ms(E')$. We claim that w.h.p. $\ms(E') \ge (1-o(1)) \ms(G)$, and this immediately implies that $\ms(T \cup R) \ge (\frac{4}{7}-o(1)) \ms(G)$. We later improve the bound to $(\frac{3}{5}-o(1)) \ms(G)$ in Subsection \ref{sec:5aug}.

To show that $\ms(E') \ge (1-o(1)) \ms(G)$, fix an arbitrary maximum matching $\match^*$ in $G$. Each edge $e \in \match^*$ is in the last $(1-1/\log n)$ fraction of the edges with the probability of $(1-1/\log n)$. Thus, $e$ is in $E'$ with the probability of $(1-1/\log n)$. Therefore, the expected number of edges from $\match^*$ that are in $E'$ is $\ms(G) (1- 1/\log n)$. By Chernoff  bound the probability that more than $2/\log n$ fraction of these edges are in the first $1/\log n$ fraction of the stream is bounded by
\begin{align*}
\exp\bigg(-\frac{|\match^*|}{3 \cdot \log n}\bigg) &\le \exp\bigg(-\frac{\sqrt{n}}{3 \log n}\bigg) \,. & \text{Since $\ms(G) \ge \sqrt{n}$.}
\end{align*}
For a large enough $n$, we have $\sqrt{n}/(3 \log n) \ge 10 \log n$. Therefore, the probability above is bounded by $\exp(-10 \log n) = n^{-10}$. Therefore, w.h.p. the size of the maximum  matching in the remaining stream is at least $\ms(G) (1- 2/\log n)= \ms(G) (1-o(1))$. 

\subsection{Finding 3-augmenting paths}

\begin{algorithm} [p]
 \KwData{A random order stream $S=\langle e_1, \ldots, e_m \rangle$ of edges of a bipartite graph $G=(A,B,E)$, and a maximal matching $\match_0$.}
 \begin{algorithmic} [1]
 \STATE Set $\tau=\frac{1}{100 \log^3 n} \,.$
 \STATE Run the following subroutines in parallel.
 \STATE \hspace{\algorithmicindent} \textbf{Subroutine 1:}
 \STATE \hspace{\algorithmicindent}\hspace{\algorithmicindent} Run $\gr$ to find a matching $P_1$ on the set of edges $(a,b) \in S_{[1,m \cdot \tau]}$ such that $a \in A(\match_0)$ and $b \notin B(\match_0)$. 
 \STATE \hspace{\algorithmicindent} \hspace{\algorithmicindent} Let $\match_P=\{ e \in \match_0 | e \cap V(P_1) \neq \O\}$ be the set of edges in $\match_0$ that are incident to the edges in $P_1$.
\STATE \hspace{\algorithmicindent} \hspace{\algorithmicindent} Run $\gr$ to find a matching $P_2$ on the set of edges $(a,b) \in S_{(m \cdot \tau, 2 m \cdot \tau]}$ such that $a \notin A(\match_0)$ and $b \in B(\match_P)$.  
 \STATE \hspace{\algorithmicindent} \textbf{Subroutine 2:}
\STATE \hspace{\algorithmicindent}\hspace{\algorithmicindent} Run $\gr$ to find a matching $Q_1$ on the set of edges $(a,b) \in S_{[1,m \cdot \tau]}$ such that $a \notin A(\match_0)$ and $b \in B(\match_0)$. 
 \STATE \hspace{\algorithmicindent} \hspace{\algorithmicindent} Let $\match_Q=\{ e \in \match_0 | e \cap V(Q_1) \neq \O\}$ be the set of edges in $\match_0$ that are incident to the edges in $Q_1$.
\STATE \hspace{\algorithmicindent} \hspace{\algorithmicindent} Run $\gr$ to find a matching $Q_2$ on the set of edges $(a,b) \in S_{(m \cdot \tau, 2m \cdot \tau]}$ such that $a \in A(\match_Q)$ and $b \notin B(\match_0)$.

\STATE \quad
\IF  {$|P_2| \ge |\match_0|/ \log n$ or $|Q_2| \ge |\match_0|/ \log n$ \tcp* {W.l.o.g., assume $|P_2| \ge |\match_0|/ \log n$.}}
\STATE Let $U$ be  maximum 3-augmenting paths for $\match_0$ using the edges $P_1$ and $P_2$.
\STATE Let $T$ be the output of $\barg$ with the input of matching $\match_0 \Delta U$, and the stream of edges $S_{(2m \cdot \tau, m]}$.
\RETURN $T \cup \match_0$.
\ELSE
 \STATE  Run the following subroutines in parallel.
  \STATE \hspace{\algorithmicindent} \textbf{Subroutine 3}: 
 \STATE \hspace{\algorithmicindent}\hspace{\algorithmicindent}  
 Store all edges  $e=(a,b) \in S_{(2m \cdot \tau, m]}$ such that $a \notin A(\match_0)$ and $b \in B(\match_P)$ and $e \cap V(P_2) =\O$. Let $R_1$ be the set of these edges.
   \STATE \hspace{\algorithmicindent} \textbf{Subroutine 4}: 
 \STATE \hspace{\algorithmicindent}\hspace{\algorithmicindent}  
 Store all edges  $e=(a,b) \in S_{(2m \cdot \tau, m]}$ such that $a \in A(\match_Q)$ and $b \notin B(\match_0)$ and $e \cap V(Q_2) =\O$. Let $R_2$ be the set of these edges.
 \STATE \hspace{\algorithmicindent} \textbf{Subroutine 5}: 
 \STATE \hspace{\algorithmicindent}\hspace{\algorithmicindent}  
 Store all edges  $e=(a,b) \in S_{(2m \cdot \tau, m]}$ such that $a \in A(\match_0)$ and $b \notin B(\match_0)$ and $e \cap V(P_1) = \O$. Let $R_3$ be the set of these edges.
 \STATE \hspace{\algorithmicindent} \textbf{Subroutine 6}: 
 \STATE \hspace{\algorithmicindent}\hspace{\algorithmicindent}  
 Store all edges  $e=(a,b) \in S_{(2m \cdot \tau, m]}$ such that $a \notin A(\match_0)$ and $b \in B(\match_0)$ and $e \cap V(Q_1) = \O$. Let $R_4$ be the set of these edges.
 
  \STATE \quad
 \STATE  Let $T$ be $\match_0 \cup P_1 \cup P_2 \cup Q_1 \cup Q_2 \cup R_1 \cup \cdots \cup R_4$.
 \RETURN $T$.
  \ENDIF 
 \end{algorithmic}
\caption{$\barg$ Algorithm for augmenting a matching using 3-augmenting paths.}
\label{alg:3arg}
\end{algorithm}

In this section we explain our algorithm for finding $3$-augmenting paths. Suppose that we are given a graph $G=(A,B,E)$ and a matching $\match_0$ in $G$. Let $R$ be the set of edges that do not interfere with any of edges in $\match_0$. Algorithm $\barg$ finds a set of $3$-augmenting paths and returns a set $T$ of the candidate edges for  augmenting $\match_0$ such that w.h.p.  $ \ms\big(T \cup R\big) \ge \big(\frac{4}{7} -o(1)\big) \ms(G)$. From the algorithm $\bm$ we know that w.h.p. the number of edges in $R$ is $\tilde{O}(n)$ and we can store all of them in the memory. Therefore, we can find the matching $\ms\big(T \cup R)$, and in fact, get a $4/7 -o(1)$ approximation of the maximum matching.

In order to demonstrate the intuition behind our algorithm, we first describe the algorithm when $\match_0$ is a maximal matching. Since $\match_0$ is maximal, we have $|\match_0| \ge \ms(G)/2$. For the sake of simplicity, we only consider the case that seems to be the worst case and we assume that the size of $|\match_0|$ is exactly $\ms(G)/2$, i.e., $|\match_0| = \ms(G)/2$. We show that Algorithm $\barg$ finds a set of $3$-augmenting paths such that by swapping the edges along these augmenting paths, we can get w.h.p. a matching $\match_1$ with the size of at least  $\big(\frac{2}{3}-o(1)\big) \ms(G)$.  Since $M_0$ is maximal and its size is $\ms(G)/2$, each edge in $M_0$ is part of a disjoint $3$-augmenting path. Therefore there exist a perfect matching between the matched vertices of part $A$ (resp., $B$) and unmatched vertices in $B$ (resp., $A$). Let use $G_U$ to denote the induced subgraph between the matched vertices in $A$ and unmatched vertices in $B$. Then, we know that the size of the maximum matching in $G_U$ is $|\match_0|$, i.e., $\ms(G_U) = |\match_0|$. 
Similarly, we define $G_L$ to be the induced subgraph graph between matched vertices in $B$ and unmatched vertices in $A$. Using a similar argument we have $\ms(G_L) = |\match_0|$. 

Simply finding a matching in $G_U$ and $G_L$ is not enough for finding $3$-augmenting paths. For example, suppose we use $\gr$ algorithm to find a matching in $G_U$ and $G_L$. Since the approximation ratio of the greedy algorithm is $1/2$, it is possible  that the greedy algorithm returns a matching of size $|\match_0|/2$ in both $G_U$ and $G_L$. The worst case scenario is when for each edge $e \in \match_0$, at most one of its endpoints is covered by these two matchings. In this case, we cannot find any $3$-augmenting path for the edges in $\match_0$. This example shows that finding $3$-augmenting paths is by nature harder than finding a matching between matched and unmatched vertices.

Suppose that we want to find a $3$-augmenting path for some edge $e \in \match_0$. Then, we should find a path $(e_U, e, e_L)$ in the graph where $e_U \in E(G_U)$ and $e_L \in E(G_L)$. We refer to $e_U$ as a \textit{upper wing} and we refer to $e_L$ as a \textit{lower wing}. Let $\tau=\frac{1}{100 \log^3 n}$ be a very small number. Algorithm \ref{alg:3arg} first finds a matching $P_1$ (respectively, $Q_1$) in $G_U$ (resp., $G_L$) using the $\tau$ fraction of the edges which is only a small portion of all edges. 
Consider an edge $e=(a,b) \in M_0$ such that $a$ is matched by $P_1$. Then, the matching edge in $P_1$ is an upper wing for $e$. If we match the other end of $e$, i.e., vertex $b$, to one of the vertices in $A(G_L)$, then we have a lower wing and we can find a $3$-augmenting path for $e$ and increase the size of the matching. Let $\match_P=\{e \in \match_0 | e\cap P_1 \neq \O\}$ be the matching edges in $\match_0$ that are incident to the edges in $P_1$. In order to find $3$-augmenting paths, we use another $\tau$ fraction of the edges to match the vertices in $B(\match_P)$ to the vertices in $A(G_L)$. Let $P_2$ be the result of $\gr$ algorithm to find this matching. We can use edges in $P_1$ and $P_2$ to find $\min\big\{|P_1|,|P_2|\big\}=|P_2|$, 3-augmenting paths for $\match_0$.

Consider the case that $|P_2|\ge|\match_0|/\log n$. In this case we can find $|\match_0|/\log n$ disjoint 3-augmenting paths. Therefore, we can increase the size of $\match_0$ by a multiplicative factor of $(1+1/\log n)$ by swapping the edges along these augmenting paths. Therefore, our algorithm can increase the size of $\match_0$ by a factor of $(1+1/\log n)$  using only $2 \tau \le \frac{1}{50\log^3 n}$ fraction of the edges which is a very small portion of all edges. In this case we increase the size of $\match_0$ and recursively call the algorithm to increase the size of the new matching. Since the size of $\match_0$ will increase by a factor of $(1+1/\log n)$ by each recursive call and initially the size of our matching is at least $\ms(G)/2$, the maximum number of recursive calls is bounded by $\log_{1+\log n} 2 \le \log n$. Eventually, our algorithm reaches a state such that $|P_2| < |\match_0|/\log n$ and at this point the algorithm has seen at the at most $2\tau\cdot \log n= \frac{1}{50 \log^2 n}$ fraction of the edges which is still very small. Therefore, we can assume that the size of $P_2$ (similarly, $Q_2$)  is at most $|\match_0|/\log n = o(|\match_0|)$.

Consider the case that the size of  $P_1$ is large. In this case, the greedy algorithm had returned a matching of size $|P_2|= o(|\match_0|)=o(|P_1|)$ between the vertices $B(\match_P)$ and $A(G_L)$ which is very small. Roughly speaking, almost all of the vertices in $B(\match_P)$ and $A(G_L)$ are remained unmatched using the greedy algorithm. Lemma \ref{lem:smallremaining}  implies that w.h.p. the number of edges between the vertices $B(\match_P)$ and $A(G_L)$ that do not interfere with the edges in $P_2$ is bounded by $\tilde{O}(n)$ and we can store all of them in the memory. Recall that there exists a perfect matching in $G_L$. Therefore, the size of the maximum matching between $B(\match_P)$ and $A(G_L)$ is $|P_1|$ and at most $2|P_2|$ edges of them interfere with the vertices in $V(P_2)$. Since we can store all edges between $B(\match_P)$ and $A(G_L)$ that are not incident to the vertices in $V(P_2)$, we can find a matching of size 
$$|P_1|-2|P_2| = |P_1| \big(1-o(1)\big) \,,$$
between $B(\match_P)$ and $A(G_L)$. Thus, we can find $|P_1|\big(1-o(1)\big)$ augmenting paths for $|\match_0|$ and increase the size of the matching. If $|P_1| \ge |\match_0|/3$, then the size of the matching returned by our algorithm is at least
\begin{align*}
|\match_0|+ |P_1| \big(1-o(1)\big) & \ge |\match_0|\big(1+1/3-o(1)\big) \\
& \ge \frac{\ms(G)}{2}\big(1+1/3-o(1)\big) & \text{Since $M_0$ is a maximal matching.} \\
& = \big(2/3-o(1)\big) \ms(G) \,,
\end{align*}
which is the desired approximation ratio. 

The remaining case is when $|P_1| < |\match_0|/3$. Recall that $P_1$ is the result of the running the greedy algorithm in order to find a matching in $G_U$. Lemma \ref{lem:smallremaining} implies that w.h.p. the number of edges in $G_U \setminus V(P_1)$ is $\tilde{O}(n)$ and we can store all of them in the memory. Let $R'$ be the set of these edges. By Lemma \ref{lem:smallgreedy} we know that 
\begin{align*}
|P_1 \cup \ms(R')| &\ge \ms(G_U)- |P_1| \\
& = |\match_0| - |P_1| \\
& \ge \frac{2|\match_0|}{3} \,. & \text{Since $|P_1| <|\match_0|/3$.} 
\end{align*}
Therefore, the algorithm can find a matching of size $2|\match_0|/3$ in $G_U$. Using a similar argument about $Q_1$, we can say that the algorithm finds a matching of size at least $2|\match_0|/3$ in $G_L$. Therefore, the number of edges in $\match_0$ that are covered by both of these matchings is at least
\begin{align*}
\frac{2|\match_0|}{3} + \frac{2|\match_0|}{3} - |\match_0| = \frac{|\match_0|}{3} \,.
\end{align*}
As the result, the algorithm can find $|\match_0|/3$ augmenting paths. By applying these augmenting paths, we get a matching of size at least
\begin{align*}
|\match_0|(1+1/3) &\ge \frac{\ms(G)}{2}(1+1/3) & \text{Since $M_0$ is a maximal matching.} \\
& = (2/3)\ms(G) \,,
\end{align*}
which is the desired approximation ratio. Roughly speaking, this shows that our algorithm has an acceptable approximation ratio when $\match_0$ is maximal. In the following theorem, we demonstrate the performance of our algorithm giving any matching $\match_0$.
\begin{theorem}
Given a bipartite graph $G=(A,B,E)$ with $n$ vertices and a matching $\match_0$. Algorithm $\barg$ returns a set of edges $T$ such that $\match_0 \subseteq T$ and w.h.p. in $n$, we have $|T| = \tilde{O}(n)$ and $\ms(T \cup R) \ge \big(\frac{4}{7}-o(1)\big) \ms(G)$ where $R=E\big(G\setminus V(\match_0)\big)$ is the set of edges in $G$ that do not interfere with the edges in $\match_0$.
\end{theorem} 
\begin{proof}
Consider an arbitrary maximum matching $\match^*$ in $G$. Let $\match^*_1$ be the set of edges in $\match^*$ that are incident to the edges in $\match_0$, and $\match^*_2 = \match^* \setminus \match^*_1$ be the set of other edges. $\match^*_1$ and $\match^*_2$ partition the edges in $\match^*$. Therefore, we have
\begin{align}
\label{eq:m1m2total}
|\match^*_1|+|\match^*_2| = |\match^*|= \ms(G) \,.
\end{align}
For every edge in $\match^*_1$, at least one of its endpoints is in $V(\match_0)$. Therefore,
\begin{align}
\label{eq:maxhalf1}
|\match^*_1| \le 2 | \match_0| \,.
\end{align}
We claim that $|\match_0|\le  |\match_1^*|$. Since edges in $\match^*_2$ do not interfere with the edges in $\match_0$, edges in $\match_0 \cup \match^*_2$ are a matching in the graph. Recall that $\match^*$ is a maximum matching in $G$. Since $\match_0 \cup \match^*_2$ is a matching we have
\begin{align*}
&|\match_0| + |\match^*_2| \le |\match^*| \\
&\Rightarrow |\match_0| + |\match^*_2| \le |\match^*_1|+|\match^*_2| & \text{By (\ref{eq:m1m2total}).} \\
&\Rightarrow |\match_0| \le |\match^*_1| \,.
\end{align*}

This along with (\ref{eq:maxhalf1}) implies that $|\match_0|= (1/2+\delta) |\match^*_1|$ where $0 \le \delta \le 1/2$. Suppose that $|\match^*_1| = \alpha \ms(G)$. Considering the matching $\match_0 \cup \ms(R)$, we have
\begin{align*}
|\match_0 \cup \ms(R)| &\ge |\match_0|+\ms(R)
\\& \ge |\match_0| + |\match^*_2| & \text{Since every edge in $\match^*_2$ is in $R$.}\\ 
& = |\match_0|+(1-\alpha) \ms(G) & \text{Since $|\match^*_1|+|\match^*_2|= \ms(G)$.}
\\ & = (1/2+\delta)|\match^*_1|+(1-\alpha) \ms(G) 
\\ & = (1/2+\delta) \alpha \ms(G)+(1-\alpha) \ms(G) & \text{Since $|\match^*_1|= \alpha \ms(G)$.}
\\ & = \ms(G) ( 1- \alpha/2 +\alpha \delta) \,.
\end{align*} 
Therefore, if $\alpha/2- \alpha \delta \le \frac{3}{7}$, we have 
$$
|\match_0 \cup \ms(R)| \ge  \ms(G) ( 1- \alpha/2 +\alpha \delta) \ge \frac{4}{7} \ms(G) \,.
$$
Since $\match_0 \subseteq T$, in the set $T$ returned by the algorithm we have $\ms(T \cup R) \ge\frac{4}{7} \ms(G)$ which is the desired bound. Therefore, for the rest of the proof we can assume that 
\begin{align}
\label{ieq:alphadelta}
\alpha/2- \alpha \delta \ge \frac{3}{7} \,.
\end{align}
 This also implies that $\alpha/2 +  \alpha \delta \le \frac{4}{7}$, since
 \begin{align}
 \label{ieq:m0bound}
 \alpha/2 +  \alpha \delta &= \alpha - (\alpha/2 - \alpha \delta) \nonumber \\
 & \le 1- (\alpha/2 - \alpha \delta) & \text{Since $\alpha \le 1$.} \nonumber \\
 & \le 4/7 \,. & \text{By (\ref{ieq:alphadelta}).}
 \end{align}

We use $G_U$ to denote the induced subgraph between the matched vertices in $A$ and unmatched vertices in $B$, i.e., $G_U$ is the induced subgraph between the vertices $A(\match_0)$ and $\overline{B(\match_0)}$. We also use $G_L$ to denote the induced subgraph between the matched vertices in $B$ and unmatched vertices in $A$. In order to find $3$-augmenting paths for the edges in $\match_0$, the algorithm should find upper wings from $G_U$ and lower wings from $G_L$. Let $\tau=\frac{1}{100 \log^3 n}$, the algorithm uses $\tau$ fraction of the edges  and runs $\gr$ algorithm to find a matching $P_1$ (respectively, $Q_1$) in $G_U$(resp., $G_L$). Let $E' \subseteq E$ be the set of edges in $E$ that are incident to the vertices in $V(\match_0)$.  $\match^*_1$ is a matching using the in $E'$. Therefore, $\ms(E') \ge |\match^*_1|$. The following claim shows that there exists a matching of size at least $(1/2-\delta) |\match^*_1|$ in both $G_U$ and $G_L$. The proof is deferred to Appendix \ref{sec:appx}. 

\begin{restatable}{claim}{matchsides}
\label{claim:matchsides}
Given a bipartite graph $G=(A,B,E)$ and a maximal matching $\match$, suppose that $|\match|=(1/2+\delta) \ms(G)$ where $0 \le \delta \le 1/2$, then there exists a matching of size at least $(1/2 - \delta)\ms(G)$ between vertices $A(\match)$ and $\overline{B(\match)}$ and also between vertices $B(\match)$ and $\overline{A(\match)}$.
\end{restatable}

The algorithm uses a small fraction of edges to find matchings $P_1$ and $Q_1$. Let $M_P$ be the edges in $M_0$ that are incident to the edges in $P_1$. Specifically, $M_P=\{e \in \match_0 | e \cap V(P_1) \neq \O \}$. Each edge in $\match_P$ is matched by the matching $P_1$. Therefore, we have an upper wing for every edge in $\match_P$. If we can find lower wings for these edges, then we have found $3$-augmenting paths for these edges, and we can increase the size of the matching. 
In order to find lower wings, the algorithm in (Subroutine 1) uses another $\tau$ fraction of the edges and calls the greedy algorithm to match vertices in $B(\match_P)$ to the vertices in $A(G_L)$. Let $P_2$ be the result of this matching. We can use edges in $P_1$ and $P_2$ to find $\min\big\{|P_1|,|P_2|\big\}=|P_2|$, 3-augmenting paths for $\match_0$. If $|P_2| \ge (1/\log n) |\match_0|$, then by switching the edges along the augmenting paths, we can increase the size of $\match_0$ by a factor of $(1+1/\log n)$. In this case, the algorithm increases the size of $\match_0$ and recursively calls itself with the new matching. Let $t$ be the number of recursive calls. Since the size of matching $\match_0$ is always bounded by $\ms(G)$, we have 
\begin{align*}
\bigg(1+\frac{1}{\log n}\bigg)^t \le \ms(G) \le n \,.
\end{align*}
Thus, $t$ is at most $\log_{1+1/\log n} n$ which is at most $2 \log^2 n$ for a large enough $n$. By the end of all recursive calls the algorithm has only used $t \cdot 2\tau \cdot m \le m/ (25 \log n)$ edges which is only a small fraction of all edges. After at most $t$ recursive calls, the algorithm reaches a state where $|P_2| < |\match_0| / \log n$. At this point the algorithm has used at most $1/(25 \log n)$ fraction of the edges.
We argue that ignoring this small fraction of the edges does not affect the size of the maximum matching. Since edges arrive in a uniformly random order, for each edge in $\match^*$ the probability that this edge is within the first $1/(25 \log n)$ fraction of the edges is $1/(25 \log n)$. Hence, the expected number of edges in $\match^*$ that are in the first $1/(25 \log n)$ fraction of the edges is $|\match^*| / (25 \log n)$. By Chernoff bound the probability that more than $2|\match^*| / (25 \log n)$ of these edges are in the first $1/(25 \log n)$ fraction of the edges is bounded by
\begin{align*}
&\exp\bigg(-\frac{|\match^*|}{3 \cdot 25 \log n}\bigg) \\
&\le \exp\bigg(-\frac{\sqrt{n}}{75 \log n}\bigg) \,. & \text{Since $\ms(G) \ge \sqrt{n}$.}
\end{align*}
For a large enough $n$, we have $\sqrt{n}/(75 \log n) \ge 10 \log n$. Therefore, the probability above is bounded by $\exp(-10 \log n) = n^{-10}$ for a large enough $n$. Therefore, w.h.p. the size of the maximum  matching in the remaining stream is at least
\begin{align*}
|\match^*| \bigg(1-\frac{2}{25 \log n}\bigg) = \ms(G) \bigg(1-\frac{2}{25 \log n}\bigg) = \ms(G) \big(1- o(1)\big) \,.
\end{align*}
Thus, the recursive calls of the algorithm does not change the size of the matching, and we can assume $|P_2| < |\match_0| / \log n$. Also, using a similar argument we can assume $|Q_2| < |\match_0| / \log n$.

By a slight abuse of notion, we suppose that $\match^*$ is a maximum matching in the remaining stream. We also suppose that $\match^*_1$ is the set of edges in $\match^*$ that are incident to the edges in $\match_0$, and $\match^*_2 = \match^* \setminus \match^*_1$ is the set of other edges in $\match^*$. Therefore,
\begin{align*}
|\match^*| = |\match^*_1| + |\match^*_2| = \ms(G) \big(1-o(1)\big) \,.
\end{align*}
Let $R_P$ be the set of edges in $\match^*_2$ that are incident to the edges in $P_1$. In other words,
\begin{align*}
R_P=\{ e \in \match^*_2 \enskip | \enskip e \cap V(P_1) \neq \O \} \,.
\end{align*}
Similarly, we define $R_Q$ to be
\begin{align*}
R_Q=\{ e \in \match^*_2 \enskip | \enskip e \cap V(Q_1) \neq \O \} \,.
\end{align*}
In order to prove the theorem, we consider two different cases on sizes of $P_1, Q_1, R_P$ and $R_Q$. The cases are as follows.
\begin{itemize}
\item \textbf{Case 1}: $|P_1|-|R_P| \ge \frac{1}{7} \ms(G)$, or $|Q_1|-|R_Q| \ge \frac{1}{7} \ms(G)$.
\vskip 0.09in
\quad Without loss of generality, we assume that $|P_1|-|R_P| \ge \ms(G)/7$. Recall that $M_P$ is the set of edges in $\match_0$ that are incident to the edges in $P_1$. Since $P_1$ is a matching in $G_U$, each edge in $M_P$ has an upper wing, and if we find lower wings for them, we have a set of $3$-augmenting path. The Algorithm in (Subroutine 1) picks a small fraction of edges and finds a greedy matching $P_2$ between the vertices in $B(\match_P)$ and $A(G_L)$. As we discussed before, we can assume that $|P_2| \le |\match_0|/\log n$. Therefore, we have
\begin{align*}
|P_2| &\le \frac{|\match_0|}{\log n} \le \frac{\ms(G)}{\log n} = o(\mu(G)) \,.
\end{align*}
Thus, the greedy algorithm has matched only a small fraction of the vertices in the induced graph between $B(\match_P)$ and $A(G_L)$. Let $G'$ be the induced subgraph between vertices $B(\match_P)$ and $A(G_L)$. By Lemma \ref{lem:smallremaining} we have that w.h.p. the number of edges in the graph $G' \setminus V(P_2)$ is bounded by $\tilde{O}(\frac{n}{\tau}) = \tilde{O}(n)$ and we can store all of them in the memory bound of our semi-streaming algorithm. The algorithm saves all of these edges in (Subroutine 3).

\quad In the following claim we show that there exists a matching in $G'$ with the size of at least $\big(|P_1| - \ms(G) \cdot 2\delta \alpha\big)$ such that its edges do not interfere with the edges in $\match^*_2$. The proof can be found in Appendix \ref{sec:appx}.
\begin{restatable}{claim}{matchminus}
\label{claim:matchminus}
Let $\match'$ be the set of edges in $\match^*_1$ that are in $G'$. Then, $|\match'|$ is at least $|P_1| - \ms(G) \cdot 2\delta \alpha$.
\end{restatable}
 We show that how this claim implies that the approximation ratio of our algorithm is at least $(4/7 -o(1))$. Suppose that $\ms(G')$ is at least $(|P_1| - \ms(G) \cdot 2\delta \alpha)$. Let $G''$ be the graph resulting from removing the vertices of $P_2$ from $G'$, i.e., $G'' = G' \setminus V(P_2)$. Since size of $P_2$ is very small and it is at most $o(\ms(G))$, we have that
\begin{align*}
\ms(G'') \ge \ms(G') - 2 |P_2| = |P_1| - \ms(G) \cdot 2\delta \alpha - o(\ms(G))= |P_1|- \ms(G)\big(2\delta \alpha +o(1)\big) \,.
\end{align*}
Since we store all of the edges in $G''$ in our memory, we can find a matching $\match'$ of size at least $|P_1|- \ms(G)\big(2\delta \alpha +o(1)\big)$ in this subgraph such that the edges of this matching do not interfere with the edges in $\match^*_2$. Each matching edge in this subgraph will be a lower wing for an edge in $M_P$. Therefore, the algorithm will find $|P_1|- \ms(G)\big(2\delta \alpha +o(1)\big)$ augmenting paths. These augmenting paths interfere with at most $|R_P|$ edges of $\match^*_2$, since none of the edges in $\match'$ interfere with the edges in $\match^*_2$. Let $T$ the union of the edges in $\match_0$ and the augmenting paths that we have found. According to what we discussed, we have
\begin{align*}
&\ms(T \cup R)\\
&\ge |M_0| + \big(|P_1|- \ms(G)\big(2\delta \alpha +o(1)\big)\big) + |\match^*_2| - |R_P| \\
&= (1/2+\delta) |\match^*_1| + |P_1|- \ms(G)\big(2\delta \alpha +o(1)\big) + |\match^*_2| - |R_P| \\
&= (1/2+\delta) \alpha \ms(G) + |P_1|- \ms(G)\big(2\delta \alpha +o(1)\big) + |\match^*_2| - |R_P| & \text{Since $|\match^*_1|= \alpha |\match^*|$.} \\
&= (1/2- \delta) \alpha \ms(G) + |P_1|- o(\ms(G)) + |\match^*_2| - |R_P| \\
&=  (1/2- \delta) \alpha \ms(G) + |P_1| + (1-\alpha) \ms(G) - |R_P| - o(\ms(G)) &\text{ Since $|\match^*_2| = |\match^*| - |\match^*_1|$.} \\
&= \big(1-\alpha/2 - \delta \alpha-  o(1) \big) \ms(G) + |P_1| - |R_P| \\
&\ge \big(1-\alpha/2 - \delta \alpha- o(1) \big) \ms(G) + \ms(G)/7 & \text{ Since $|P_1|-|R_P| \ge \ms(G)/7$.} \\
&= \big(8/7-\alpha/2 -\delta \alpha - o(1) \big) \ms(G) \,.
\end{align*}
By inequality above and (\ref{ieq:m0bound}), we have,
\begin{align*}
\ms(T \cup R) &\ge \big(8/7-\alpha/2 -\delta \alpha - o(1) \big) \ms(G) \\
&\ge \big(4/7- o(1) \big) \ms(G) \,, & \text{By (\ref{ieq:m0bound}).} \\
\end{align*}
which is our desired bound. We can use a similar argument to show that when $|Q_1|-|R_Q| \ge \frac{1}{7} \ms(G)$, the approximation ratio of algorithm is at least $\big(\frac{4}{7}-o(1)\big)$.
\item \textbf{Case 2}: $|P_1|-|R_P| < \frac{1}{7} \ms(G)$, and $|Q_1|-|R_Q| < \frac{1}{7} \ms(G)$. 
\vskip 0.09in
\quad The remaining case is when $|P_1|-|R_P| < \frac{1}{7} \ms(G)$ and $|Q_1|-|R_Q| < \frac{1}{7} \ms(G)$. Recall that $P_1$ and $Q_1$ are the result of the greedy matching algorithm using $\tau$ fraction of the edges in $G_U$ and $G_L$. Therefore, by Lemma \ref{lem:smallremaining}, we know that with w.h.p. we can store all of the edges in $G_U \setminus V(P_1)$ and also all the edges in $G_L \setminus V(Q_1)$ within a memory of size $\tilde{O}(n/\tau) = \tilde{O}(n)$ which is within the memory bound of the semi-streaming algorithms. The algorithm stores all the edges in $G_U \setminus V(P_1)$ and also all the edges in $G_L \setminus V(Q_1)$ in (Subroutine 5) and (Subroutine 6) respectively. Let $R_3$ and $R_4$ be the set of these edges respectively, and $T$ be $P_1 \cup Q_1 \cup R_3 \cup R_4$. We show that $\ms(T) \ge \big(\frac{4}{7} - o(1)\big) \ms(G)$.

\quad We claim that using the edges in $R_3$ and $P_1$ we can find a matching of size of $(\frac{2}{7} - o(1)) \ms(G)$ in $G_U$. Consider the matching $\match^*_1$.  By Claim \ref{claim:matchsides} we know there exists a matching of size at least $(1/2-\delta)|\match^*_1|$ in $G_U$. Let $\match$ be this matching in $G_U$, and $\match'$ be the set of the edges in $\match$ that interfere with the edges in $P_1$, i.e., $\match' = \{ e \in \match \enskip | \enskip e \cap V(P_1) \neq \O \}$. We show that the size of $\match'$ is at most $2|P_1| - |R_P|$. The total number of vertices covered by $P_1$ is exactly $2|P_1|$. However, we know that $|R_P|$ edges from $P_1$ cover at least one vertex from $\match^*_2$. Since vertices of $\match^*_1$ and $\match^*_2$ are disjoint, it implies that the number of edges in $\match^*_1$ that are covered by $P_1$ is at most $2|P_1|-|R_P|$. Thus, $|\match'| \le 2|P_1| - |R_P|$. Note that edges in $\match \setminus \match'$ do not share a vertex with the edges in $P_1$. Therefore, the size of maximum matching in $R_3 \cup P_1$ is at least,
\begin{align*}
|P_1| + |\match| - |\match'| &\ge |P_1| + |\match| - (2|P_1| - |R_P|) &\text{Since $|\match'| \le 2|P_1| - |R_P|$.} \\
&=|R_P|-|P_1| +|\match| \\
&> |\match| - \ms(G)/7 \,. & \text{By the assumption of Case 2.}
\end{align*}
Since the size $\match$ is at least $(1/2-\delta)|\match^*_1|$, we have
\begin{align*}
\ms(R_3 \cup P_1) &\ge |\match| -  \ms(G)/7 \\
& \ge (1/2-\delta)|\match^*_1| - \ms(G)/7 \\
& = (1/2-\delta)\alpha |\match^*| -  \ms(G)/7 &\text{Since $|\match^*_1|= \alpha |\match^*|$.} \\
& = (1/2-\delta - o(1))\alpha \ms(G) - \ms(G)/7 \\
& = \ms(G) ( \alpha/2 - \alpha \delta -1/7 -o(1)) \\
& \ge \ms(G) ( 3/7 -1/7 -o(1)) & \text{By (\ref{ieq:alphadelta}).} \\
& = \ms(G) ( 2/7 -o(1)) \,.
\end{align*}
Using a similar argument for $R_4 \cup Q_1$, we can say that $\ms(R_4 \cup Q_1) \ge (\frac{2}{7} - o(1)) \ms(G)$. Therefore,
\begin{align*}
\ms(T) = \ms(P_1 \cup Q_1 \cup R_3 \cup R_4)= \ms(P_1 \cup R_3) + \ms( Q_1 \cup R_4) \ge \ms(G) ( 4/7 -o(1)) \,.
\end{align*}
Which is our desired bound. This completes the analysis for the only remaining case and completes the proof of the theorem.
\end{itemize}

\end{proof}
\subsection{Finding 5-augmenting paths}
\label{sec:5aug}
In this section, we slightly modify $\barg$ algorithm to find a small set of $5$-augmenting paths. The algorithm receives an initial matching $\match_0$, and finds a set of augmenting paths for $\match_0$. We use $G_U$ to denote the induced subgraph between the matched vertices in $A$ and unmatched vertices in $B$, i.e., $G_U$ is the induced subgraph between the vertices $A(\match_0)$ and $\overline{B(\match_0)}$. We also use $G_L$ to denote the induced subgraph between the matched vertices in $B$ and unmatched vertices in $A$. In order to find $3$-augmenting paths for the edges in $\match_0$, the algorithm should find upper wings from $G_U$ and lower wings from $G_L$. Similar to our algorithm in the previous section, the algorithm uses a very small fraction of the edges to find a matching $P_1$ in $G_U$ and $Q_1$ in $G_L$. Let $\match_P$ and $\match_Q$ be the matching edges in $\match_0$ that are incident to the edges in $P_1$ and $Q_1$ respectively.

Consider an edge $e \in \match_P \cap \match_Q$, we can find an upper wing for this edge using the edges in $P_1$, and we can also find a lower wing for this edge using the edges in $Q_1$. Therefore we can find $|\match_P \cap \match_Q|$ disjoint 3-augmenting paths. If the size of $\match_P \cap \match_Q$ is large, e.g., larger than $(1/\log n) |\match_0|$, our algorithm has found a large set of 3-augmenting paths by only using $2 \tau$ fraction of the edges which is very small. We can then switch the edges along these augmenting paths, and recursively find augmenting paths for the new matching. Therefore we can assume the size of $\match_P \cap \match_Q$ is at most $o(|\match_0|)$. In other words we can assume the sets $\match_P$ and $\match_Q$ are almost disjoint.

Our algorithm (formally as Algorithm \ref{alg:5arg}) then finds a matching between $B(\match_P)$ and $A(\match_Q)$. Consider an edge $e$ between $B(\match_P)$ and $A(\match_Q)$, this edge connects two matching edges $e_0 \in \match_P$ and $e_1 \in \match_Q$ to each other. Recall using the edges in $P_1$, we have a upper wing for $e_0$. Let $e_U$ be the upper wing. Also using the edges in $Q_1$, we have a lower wing for $e_1$. Let $e_L$ be the lower wing. Then $(e_U, e_0, e, e_1, e_L)$ is a 5-augmenting path in our graph. Our algorithm uses a very small fraction of the edges to find a matching between $B(\match_P)$ and $A(\match_Q)$. If it finds a matching with a large size, then we have a large number of $5$-augmenting paths. We can exchange the edges along these augmenting paths and recursively solve the problem for the new matching. Otherwise, we can assume that the size of the matching found between $B(\match_P)$ and $A(\match_Q)$ is very small. In that case Lemma \ref{lem:smallremaining} roughly implies that we can store all edges between $B(\match_P)$ and $A(\match_Q)$ using a very small memory. We show that our algorithm finds $\big(\frac{3}{5} - o(1)\big)$ approximation of the maximum matching.

\begin{algorithm} [p]
 \KwData{A random order stream $S=\langle e_1, \ldots, e_m \rangle$ of edges of a bipartite graph $G=(A,B,E)$, and a maximal matching $\match_0$.}
 \begin{algorithmic} [1]
 \STATE Set $\tau=\frac{1}{100 \log^3 n} \,.$
 \STATE Run the following subroutines in parallel.
 \STATE \hspace{\algorithmicindent} \textbf{Subroutine 1:}
 \STATE \hspace{\algorithmicindent}\hspace{\algorithmicindent} Run $\gr$ to find a matching $P_1$ on the set of edges $(a,b) \in S_{[1,m \cdot \tau]}$ such that $a \in A(\match_0)$ and $b \notin B(\match_0)$. 
 \STATE \hspace{\algorithmicindent} \hspace{\algorithmicindent} Let $\match_P=\{ e \in \match_0 | e \cap V(P_1) \neq \O\}$ be the set of edges in $\match_0$ that are incident to the edges in $P_1$.
\STATE \hspace{\algorithmicindent} \hspace{\algorithmicindent} Run $\gr$ to find a matching $P_2$ on the set of edges $(a,b) \in S_{(m \cdot \tau, 2 m \cdot \tau]}$ such that $a \notin A(\match_0)$ and $b \in B(\match_P)$.  
 \STATE \hspace{\algorithmicindent} \textbf{Subroutine 2:}
\STATE \hspace{\algorithmicindent}\hspace{\algorithmicindent} Run $\gr$ to find a matching $Q_1$ on the set of edges $(a,b) \in S_{[1,m \cdot \tau]}$ such that $a \notin A(\match_0)$ and $b \in B(\match_0)$. 
 \STATE \hspace{\algorithmicindent} \hspace{\algorithmicindent} Let $\match_Q=\{ e \in \match_0 | e \cap V(Q_1) \neq \O\}$ be the set of edges in $\match_0$ that are incident to the edges in $Q_1$.
\STATE \hspace{\algorithmicindent} \hspace{\algorithmicindent} Run $\gr$ to find a matching $Q_2$ on the set of edges $(a,b) \in S_{(m \cdot \tau, 2m \cdot \tau]}$ such that $a \in A(\match_Q)$ and $b \notin B(\match_0)$.
\STATE \quad

\STATE Run $\gr$ to find a matching $C$ on the set of edges $(a,b) \in S_{(2m \cdot \tau,3m \cdot \tau]}$ such that $a \in A(\match_Q)$ and $b \in B(\match_P)$. 

\STATE Let $T'= P_1 \cup P_2 \cup Q_1 \cup Q_2 \cup C$.
\IF  {there are at least $|M_0|/\log n$ augmenting paths using the edges in $T'$}
\STATE Let $U$ be  maximum augmenting paths for $\match_0$ using the edges $T'$.
\STATE Let $T$ be the output of $\farg$ with the input of matching $\match_0 \Delta U$, and the stream of edges $S_{(3m \cdot \tau, m]}$.
\RETURN $T \cup \match_0$.
\ELSE
 \STATE  Run the following subroutines in parallel.
  \STATE \hspace{\algorithmicindent} \textbf{Subroutine 3}: 
 \STATE \hspace{\algorithmicindent}\hspace{\algorithmicindent}  
 Store all edges  $e=(a,b) \in S_{(3m \cdot \tau, m]}$ such that $a \notin A(\match_0)$ and $b \in B(\match_P)$ and $e \cap V(P_2) =\O$. Let $R_1$ be the set of these edges.
   \STATE \hspace{\algorithmicindent} \textbf{Subroutine 4}: 
 \STATE \hspace{\algorithmicindent}\hspace{\algorithmicindent}  
 Store all edges  $e=(a,b) \in S_{(3m \cdot \tau, m]}$ such that $a \in A(\match_Q)$ and $b \notin B(\match_0)$ and $e \cap V(Q_2) =\O$. Let $R_2$ be the set of these edges.
 \STATE \hspace{\algorithmicindent} \textbf{Subroutine 5}: 
 \STATE \hspace{\algorithmicindent}\hspace{\algorithmicindent}  
 Store all edges  $e=(a,b) \in S_{(3m \cdot \tau, m]}$ such that $a \in A(\match_0)$ and $b \notin B(\match_0)$ and $e \cap V(P_1) = \O$. Let $R_3$ be the set of these edges.
 \STATE \hspace{\algorithmicindent} \textbf{Subroutine 6}: 
 \STATE \hspace{\algorithmicindent}\hspace{\algorithmicindent}  
 Store all edges  $e=(a,b) \in S_{(3m \cdot \tau, m]}$ such that $a \notin A(\match_0)$ and $b \in B(\match_0)$ and $e \cap V(Q_1) = \O$. Let $R_4$ be the set of these edges.
 
  \STATE \hspace{\algorithmicindent} \textbf{Subroutine 7}: 
 \STATE \hspace{\algorithmicindent}\hspace{\algorithmicindent}  
 Store all edges  $e=(a,b) \in S_{(3m \cdot \tau, m]}$ such that $a \in A(\match_Q)$ and $b \in B(\match_P)$ and $e \cap V(C) = \O$. Let $R_5$ be the set of these edges.
 
  \STATE \quad
 \STATE  Let $T$ be $\match_0 \cup P_1 \cup P_2 \cup Q_1 \cup Q_2 \cup C \cup R_1 \cup \cdots \cup R_5$.
 \RETURN $T$.
  \ENDIF 
 \end{algorithmic}
\caption{$\farg$ Algorithm for augmenting a matching using 3 or 5 augmenting paths.}
\label{alg:5arg}
\end{algorithm}

\begin{theorem}
Given a bipartite graph $G=(A,B,E)$ with $n$ vertices and a matching $\match_0$. Algorithm $\farg$ returns a set of edges $T$ such that $\match_0 \subseteq T$ and w.h.p. in $n$, we have $|T| = \tilde{O}(n)$ and $\ms(T \cup R) \ge \big(\frac{3}{5}-o(1)\big) \ms(G)$ where $R=E\big(G\setminus V(\match_0)\big)$ is the set of edges in $G$ that do not interfere with the edges in $\match_0$.
\end{theorem}
\begin{proof}
Consider an arbitrary maximum matching $\match^*$ in $G$. Let $\match^*_1$ be the set of edges in $\match^*$ that are incident to the edges in $\match_0$, and $\match^*_2 = \match^* \setminus \match^*_1$ be the set of other edges. $\match^*_1$ and $\match^*_2$ partition the edges in $\match^*$. Therefore, we have
$|\match^*_1|+|\match^*_2| = |\match^*|= \ms(G)$.
For every edge in $\match^*_1$, at least one of its endpoints is in $V(\match_0)$. Therefore,
\begin{align*}
|\match_0| \le |\match^*_1| \le 2 | \match_0| \,.
\end{align*}

Let's suppose that $|\match_0|= (1/2+\delta) |\match^*_1|$ where $0 \le \delta \le 1/2$. Also, Suppose that $|\match^*_1| = \alpha |\match^*|$. Considering the matching $\match_0 \cup \ms(R)$, we have
\begin{align*}
|\match_0 \cup \ms(R)| &\ge |\match_0|+\ms(R)
\\& \ge |\match_0| + |\match^*_2| & \text{Since every edge in $\match^*_2$ is in $R$.}\\ 
& = |\match_0|+(1-\alpha) \ms(G) & \text{Since $|\match^*_1|+|\match^*_2|= \ms(G)$.}
\\ & = (1/2+\delta)|\match^*_1|+(1-\alpha) \ms(G) 
\\ & = (1/2+\delta) \alpha \ms(G)+(1-\alpha) \ms(G) & \text{Since $|\match^*_1|= \alpha \ms(G)$.}
\\ & = \ms(G) ( 1- \alpha/2 +\alpha \delta) \,.
\end{align*} 
Therefore, if $\alpha/2- \alpha \delta \le \frac{2}{5}$, we have 
$$
|\match_0 \cup \ms(R)| \ge  \ms(G) ( 1- \alpha/2 +\alpha \delta) \ge \frac{3}{5} \ms(G) \,.
$$
The algorithm returns a set $T$ such that $\match_0 \subseteq T$. Therefore, in this case we have $\ms(T \cup R) \ge\frac{3}{5} \ms(G)$ which is the desired bound. Therefore, for the rest of the proof we assume that 
\begin{align}
\label{ieq:alphadelta-2}
\alpha/2- \alpha \delta \ge \frac{2}{5} \,.
\end{align}

We use $G_U$ to denote the induced subgraph between the matched vertices in $A$ and unmatched vertices in $B$, i.e., $G_U$ is the induced subgraph between the vertices $A(\match_0)$ and $\overline{B(\match_0)}$. We also use $G_L$ to denote the induced subgraph between the matched vertices in $B$ and unmatched vertices in $A$. In order to find $3$-augmenting paths for the edges in $\match_0$, the algorithm should find upper wings from $G_U$ and lower wings from $G_L$. Let $\tau=\frac{1}{100 \log^3 n}$, the algorithm uses $\tau$ fraction of the edges  and runs $\gr$ algorithm to find a matching $P_1$ (respectively, $Q_1$) in $G_U$(resp., $G_L$). Let $E' \subseteq E$ be the set of edges in $E$ that are incident to the vertices in $V(\match_0)$.  Matching $\match_0$ is maximal among the edges in $E'$. Also, $\match^*_1$ is a matching in $E'$. Therefore, $\ms(E') \ge |\match^*_1|$. By Claim \ref{claim:matchsides} there exists a matching of size $(1/2-\delta) |\match^*_1|$ in both $G_U$ and $G_L$. Let $M_P$ be the edges in $M_0$ that are incident to the edges in $P_1$. In other words, $\match_P=\{e \in \match_0 | e \cap V(P_1) \neq \O \}$. Similarly, we define $\match_Q$ to be $\match_Q=\{e \in \match_0 | e \cap V(Q_1) \neq \O \}$.  Each edge in $\match_P$ is matched by the matching $P_1$. Therefore, we have an upper wing for every edge in $\match_P$. If we can find lower wings for these edges, then we have found $3$-augmenting paths for these edges, and we can increase the size of the matching. 

In order to find lower wings, the algorithm in (Subroutine 1) uses another $\tau$ fraction of the edges and calls the greedy algorithm to match vertices in $B(\match_P)$ to the vertices in $A(G_L)$. Let $P_2$ be the result of this matching. We can use edges in $P_1$ and $P_2$ to find $\min\big\{|P_1|,|P_2|\big\}=|P_2|$, 3-augmenting paths for $\match_0$. If $|P_2| \ge (1/\log n) |\match_0|$, then by switching the edges along the augmenting paths, we can increase the size of $\match_0$ by the factor of $(1+1/\log n)$. In this case, the algorithm increases the size of $\match_0$ and recursively calls itself to find augmenting paths for the new matching. Eventually, the algorithm reaches a state such that it cannot increase the size of matching using the edges in $P_1$ and $P_2$. We then have $|P_2| < (1/\log n) |\match_0|$. The algorithm also finds a matching $Q_2$ between $A(\match_Q)$ and $B(G_U)$. Using a similar argument for $Q_1$ and $Q_2$, we can say that the algorithm eventually reaches a state that we have $|Q_2| < (1/\log n) |\match_0|$. Also, consider an edge in $\match_P \cap \match_Q$, for these edges we have both upper and lower wings. Therefore, we have augmenting paths for these edges. If the size of $\match_P \cap \match_Q$ is large, we can increase the size of matching $\match_0$ by these augmenting paths, and recursively find augmenting paths for the new matching. Thus, the algorithm eventually reaches a state such that we have $|\match_P \cap \match_Q| < (1/\log n) |\match_0|$.

Our algorithm then uses another $\tau$ fraction of the edges to find a matching $C$ between vertices $B(\match_P)$ and $A(\match_Q)$. We claim that we can find $|C|$ augmenting paths using these edges. Consider an edge $e \in C$, this edge is between $B(\match_P)$ and $A(\match_Q)$, and it connects two matching edges $e_0 \in \match_P$ and $e_1 \in \match_Q$ to each other. If $e_0 = e_1$, then we already have both upper and lower wings for $e_0$. Therefore, we have an augmenting path for $e_0$. Consider the case that $e_0 \neq e_1$. Recall that using the edges in $P_1$, we have a upper wing for $e_0$. Let $e_U$ be the upper wing. Also using the edges in $Q_1$, we have a lower wing for $e_1$. Let $e_L$ be the lower wing. Then $(e_U, e_0, e, e_1, e_L)$ is a 5-augmenting path in our graph. Therefore we have $|C|$ augmenting paths. If the size of $C$ is large, we can increase the size of matching $\match_0$ by these augmenting paths, and recursively find augmenting paths for the new matching. Thus, the algorithm eventually reaches a state such that we have $|C| < (1/\log n) |\match_0|$.

Let $t$ be the number of recursive calls in our algorithm. In each recursive call the algorithm increases the size of $\match_0$ by a multiplicative factor of $(1+1/\log n)$. Therefore, we have $t \le \log_{1+1/\log n} n$ which is at most $2 \log^2 n$ for a large enough $n$. 
Before each recursive call, the algorithm uses at most $3\tau$ fraction of the edges. Therefore, by the end of all recursive calls the algorithm has only used $t \cdot 3\tau \cdot m \le m/ (16 \log n)$ edges which is only a small fraction of all edges. Since edges arrive in a uniformly random order, for each edge in $\match^*$ the probability that this edge is within the first $1/(16 \log n)$ fraction of the edges is $1/(16 \log n)$. Hence, the expected number of edges in $\match^*$ that are in the first $1/(16 \log n)$ fraction of the edges is $|\match^*| / (16 \log n)$. Using Chernoff bound we can show that the probability that more than $|\match^*| / (8 \log n)$ of these edges are in the first $1/(16 \log n)$ fraction of the edges is bounded by $n^{-10}$ for a large enough $n$. Therefore, w.h.p. the size of the maximum  matching in the remaining stream is at least
\begin{align*}
|\match^*| \bigg(1-\frac{1}{8 \log n}\bigg) = \ms(G) \bigg(1-\frac{1}{8 \log n}\bigg) = \ms(G) \big(1- o(1)\big) \,.
\end{align*}
Recall that by the end of all recursive calls the size of each of 
$P_2, Q_2, (\match_P \cap \match_Q)$ and $C$ is bounded by $|M_0|/\log n$.

By a slight abuse of notion, we suppose that $\match^*$ is a maximum matching in the remaining stream. We also suppose that $\match^*_1$ is the set of edges in $\match^*$ that are incident to the edges in $\match_0$, and $\match^*_2 = \match^* \setminus \match^*_1$ is the set of other edges in $\match^*$. Therefore,
\begin{align*}
|\match^*|= |\match^*_1| + |\match^*_2| = \ms(G) \big(1-o(1)\big) \,.
\end{align*}
Let $R_P$ be the set of edges in $\match^*_2$ that are incident to the edges in $P_1$. In other words,
\begin{align*}
R_P=\{ e \in \match^*_2 \enskip | \enskip e \cap V(P_1) \neq \O \} \,.
\end{align*}
Similarly, we define $R_Q$ to be
\begin{align*}
R_Q=\{ e \in \match^*_2 \enskip | \enskip e \cap V(Q_1) \neq \O \} \,.
\end{align*}
Also, let $\match^*_C$ to be the edges in $\match^*$ that are between vertices $A(\match_Q)$ and $B(\match_P)$, i.e.,
\begin{align*}
\match^*_C = \{ (a,b) \in \match^* \enskip | \enskip a \in A(\match_Q), b \in B(\match_P) \} \,. 
\end{align*}
Then, all edges in $\match^*_C$ are also in $\match^*_1$.

In order to prove the theorem, we consider the following cases. 

\begin{itemize}
\item \textbf{Case 1}: $|\match^*_C| \ge \big (1/5 - 2 \delta \alpha \big)\mu(G)$. 
\vskip 0.09in
\quad Every edge in $\match^*_C$ is between vertices $A(\match_Q)$ and $B(\match_P)$. As we previously discussed, every edge in $\match^*_C$ forms an augmenting path. We show that our algorithm finds a large number of these edges. The algorithm first uses a small fraction of the edges to find a matching $C$ between vertices $A(\match_Q)$ and $B(\match_P)$. According to what we discussed, we can assume $|C| \le |\match_0|/\log n$. By Lemma \ref{lem:smallremaining}, we know that with w.h.p. we can store all of the edges between vertices $A(\match_Q)$ and $B(\match_P)$ that are not incident to the edges in $C$ within a memory of size $\tilde{O}(n/\tau) = \tilde{O}(n)$. The algorithm stores all these edges in (Subroutine 7). Let $R_5$ be the set of these edges, and $T$ be $P_1 \cup Q_1 \cup R_5$. We claim that $\mu(T \cup \match_0) \ge (\frac{3}{5} - o(1)) \ms(G)$.

We first show that $\mu(R_5) \ge |\match^*_C| - o\big(\mu(G)\big)$. Set $R_5$ contains all edges between vertices $A(\match_Q)$ and $B(\match_P)$ that are not incident to the edges in $C$. Since there are at most $2|C|$ edges in $\match^*_C$ that are incident to a vertex in $V(C)$, we have
\begin{align}
\label{ieq:matchr5}
\mu(R_5) &\ge |\match^*_C| - 2|C| \nonumber \\
&\ge |\match^*_C| - 2 |\match_0| /\log n \nonumber \\
&\ge |\match^*_C| - 2 \mu(G)/\log n \nonumber \\
&= |\match^*_C| - o\big(\mu(G)\big) \,.
\end{align}
Each edge in $R_5$ is between vertices $A(\match_Q)$ and $B(\match_P)$. According to what we previously discussed, each edge in $R_5$ along with the edges in $P_1$ and $Q_1$ forms $3$ or $5$ augmenting paths for $\match_0$. Therefore, we can find $\mu(R_5)$ disjoint augmenting paths. It follows that the size of the matching in $T\cup \match_0$ is at least
\begin{align*}
&\mu(T \cup \match_0)\\
 &\ge  |\match_0| + \mu(R_5)  \\
&\ge |\match_0| + |\match^*_C| - o\big(\mu(G)\big) & \text{By (\ref{ieq:matchr5}).}\\
&\ge |\match_0| + \big (1/5 - 2 \delta \alpha -o(1)\big)\mu(G) & \text{By the assumption of Case 1.}\\
&= (1/2+ \delta)|\match^*_1| + \big (1/5 - 2 \delta \alpha -o(1)\big)\mu(G) & \text{Since $|\match_0| = (1/2+ \delta)|\match^*_1|$.} \\
&=  \big(1/2+ \delta- o(1)\big)\alpha\mu(G) + \big (1/5 - 2 \delta \alpha -o(1)\big)\mu(G) & \text{Since $|\match^*_1| = \alpha|\match^*|$.} \\
&= \big(1/5+ \alpha/2 - \delta \alpha - o(1)\big) \mu(G) \\
&\ge \big(3/5-o(1)\big)\ms(G) \,, & \text{By (\ref{ieq:alphadelta-2}).}
\end{align*}
which is the desired bound, and it completes the proof for this case.
\item \textbf{Case 2}: $|P_1|-|R_P| + |Q_1|- |Q_P| \le \big(\alpha - 2 \alpha \delta - 3/5\big) \mu(G)$. 
\vskip 0.09in
\quad  Recall that $P_1$ and $Q_1$ are the result of the greedy matching algorithm using $\tau$ fraction of the edges in $G_U$ and $G_L$. Therefore, by Lemma \ref{lem:smallremaining}, we know that with w.h.p. we can store all of the edges in $G_U \setminus V(P_1)$ and also all the edges in $G_L \setminus V(Q_1)$ within a memory of size $\tilde{O}(n/\tau) = \tilde{O}(n)$ which is within the memory bound of the semi-streaming algorithms. The algorithm stores all the edges in $G_U \setminus V(P_1)$ and also all the edges in $G_L \setminus V(Q_1)$ in (Subroutine 5) and (Subroutine 6) respectively. Let $R_3$ and $R_4$ be the set of these edges respectively, and $T$ be $P_1 \cup Q_1 \cup R_3 \cup R_4$. We show that $\ms(T) \ge (\frac{3}{5} - o(1)) \ms(G)$.

\quad By Claim \ref{claim:matchsides} we know there exists a matching of size at least $(1/2-\delta)|\match^*_1|$ in both $G_U$ and $G_L$ using the edges in $\match^*_1$. Therefore, there exists a matching of size at least $(1-2\delta)|\match^*_1|$ in $G_U \cup G_L$. Let $\match$ be this matching, and $\match'$ be the set of the edges in $\match$ that interfere with the edges in $P_1$ or $Q_1$, i.e., $\match' = \{ e \in \match \enskip | \enskip e \cap \big(V(P_1) \cup V(Q_1)\big) \neq \O \}$. We show that the size of $\match'$ is at most $2|P_1| + 2|Q_1| - |R_P| - |R_Q|$. The total number of vertices covered by $P_1$ and $Q_1$ is exactly $2|P_1|+2|Q_1|$. However, we know that $|R_P|$ edges from $P_1$ cover at least one vertex from $\match^*_2$. Since vertices of $\match^*_1$ and $\match^*_2$ are disjoint, it implies that the number of edges from $\match^*_1$ that are covered by $P_1$  is at most $2|P_1|-|R_P|$. Using a similar argument, the number of edges from $\match^*_1$ that are covered by $Q_1$  is at most $2|Q_1|-|R_Q|$. Thus, 
\begin{align}
\label{ieq:sizemp}
|\match'| \le 2|P_1| + 2|Q_1| - |R_P| - |R_Q| \,.
\end{align} 
 Note that edges in $\match \setminus \match'$ do not share a vertex with the edges in $P_1$ or $Q_1$. Therefore, the size of maximum matching in $ P_1 \cup Q_1 \cup R_3 \cup R_4$ is at least
\begin{align*}
&|P_1| +|Q_1| + |\match| - |\match'| \\
&\ge |P_1| +|Q_1|+ |\match| - (2|P_1| + 2|Q_1| - |R_P| - |R_Q|) &\text{By (\ref{ieq:sizemp}).} \\
&=|R_P|+|R_Q|-|P_1|-|Q_1| +|\match| \\
&\ge |\match| - \big(\alpha - 2 \alpha \delta - 3/5\big) \mu(G) \,. & \text{By the assumption of Case 2.}
\end{align*}
Since the size $\match$ is at least $(1-2\delta)|\match^*_1|$, we have
\begin{align*}
\ms(T) &\ge |\match| - \big(\alpha - 2 \alpha \delta - 3/5\big) \mu(G) \\
& \ge (1-2\delta)|\match^*_1| - \big(\alpha - 2 \alpha \delta - 3/5\big) \mu(G) \\
& = (1-2\delta)\alpha |\match^*| - \big(\alpha - 2 \alpha \delta - 3/5\big) \mu(G) \\
& = (1-2\delta - o(1))\alpha \ms(G) - \big(\alpha - 2 \alpha \delta - 3/5\big) \mu(G) \\
& = \ms(G) \big( 3/5  -o(1)\big)\,.
\end{align*}
Which is our desired bound. This completes the analysis for this case.
\item \textbf{Case 3}: $|\match^*_C| < \big (1/5 - 2 \delta \alpha \big)\mu(G)$ and $|P_1|-|R_P| + |Q_1|- |Q_P| > \big(\alpha - 2 \alpha \delta - 3/5\big) \mu(G)$.
\vskip 0.09in
\quad The only remaining case is when $|\match^*_C| < \big (1/5 - 2 \delta \alpha \big)\mu(G)$ and $|P_1|-|R_P| + |Q_1|- |Q_P| > \big(\alpha - 2 \alpha \delta - 3/5\big) \mu(G)$. Recall that $\match_P$ is the set of edges in $\match_0$ that are incident to the edges in $P_1$. Since $P_1$ is a matching in $G_U$, each edge in $M_P$ has an upper wing, and if we find lower wings for them, we have a set of $3$-augmenting path. The Algorithm in (Subroutine 1) picks a small fraction of edges and finds a greedy matching $P_2$ between the vertices in $B(\match_P)$ and $A(G_L)$. As we discussed before, we can assume that $|P_2| \le |\match_0|/\log n$. Thus, we have $|P_2|= o(\mu(G))$.
Therefore, the greedy algorithm has matched only a small fraction of the vertices in the induced graph between $B(\match_P)$ and $A(G_L)$. Let $G_P$ be the induced subgraph between vertices $B(\match_P)$ and $A(G_L)$. By Lemma \ref{lem:smallremaining} we have that w.h.p. the number of edges in the graph $G_P \setminus V(P_2)$ is bounded by $\tilde{O}(\frac{n}{\tau}) = \tilde{O}(n)$ and we can store all of them in the memory bound of our semi-streaming algorithm. The algorithm saves all of these edges in (Subroutine 3). Let $R_1$ be the set of these edges. Similarly, let  $G_Q$ be the induced subgraph between vertices $A(\match_Q)$ and $B(G_U)$.   By Lemma \ref{lem:smallremaining} we have that w.h.p. the number of edges in the graph $G_Q \setminus V(Q_2)$ is bounded by $\tilde{O}(\frac{n}{\tau}) = \tilde{O}(n)$ and we can store all of them in the memory bound of our semi-streaming algorithm. Let $R_2$ be the set of these edges.

We claim that $\mu(G_P) + \mu(G_Q)$ is at least $|P_1|+|Q_1|- \mu(G)\big(1/5+o(1)\big)$. Every edge of $\match^*_1$ is covered by the vertices of $\match_0$. Recall that set $\match^*_C$ is the set of edges in $\match^*_1$ that are between vertices $A(\match_Q)$ and $B(\match_P)$. Let $V'=V(\match_0) \setminus \big(A(\match_Q) \cup B(\match_P)\big)$. It follows that the number of edges in $\match^*_1$ that are not covered by the vertices in $V'$ is bounded by
\begin{align*}
\mu(G_P)+\mu(G_Q) + |\match^*_C| \,.
\end{align*}
Therefore we should have
\begin{align*}
|\match^*_1| &\le |V'| + \mu(G_P)+\mu(G_Q) + |\match^*_C| \\
& = 2|\match_0| - |P_1| - |Q_1| + \mu(G_P)+\mu(G_Q) + |\match^*_C| \,. 
\end{align*}
Therefore,
\begin{align}
\label{ieq:mupq}
&\mu(G_P)+\mu(G_Q) \nonumber \\
&\ge |P_1| + |Q_1| + |\match^*_1| - 2|\match_0|  - |\match^*_C| \nonumber \\
&= |P_1| + |Q_1| + \big(\alpha - o(1)\big) \mu(G) - 2|\match_0| - |\match^*_C|& \text{Since $|\match^*_1|= \alpha |\match^*|$.} \nonumber \\
&= |P_1| + |Q_1| - \big(2\delta\alpha + o(1)\big) \mu(G) - |\match^*_C|& \text{Since $|\match_0|= (1/2+\delta) |\match^*_1|$.} \nonumber \\
&\ge |P_1| + |Q_1| - \mu(G)\big(1/5+o(1)\big) \,. &\text{Since $|\match^*_C| < \big (1/5 - 2 \delta \alpha \big)\mu(G)$.}
\end{align}
The algorithm stores in sets $R_1$ and $R_2$ all edges in $G_P$ and $G_Q$ that do not interfere with the edges in $P_2$ and $Q_2$. Since the size of $P_2$ and $Q_2$ is at most $o(\mu(G))$, we have
\begin{align}
\label{ieq:mur1r2}
\mu(R_1) + \mu(R_2)  &\ge \mu(G_P)+\mu(G_Q) - 2|P_2| - 2|Q_2| \nonumber \\
&=\mu(G_P)+\mu(G_Q) - o(\mu(G)) \nonumber \\
&\ge  |P_1| + |Q_1| - \mu(G)\big(1/5+o(1)\big) \,. &\text{By (\ref{ieq:mupq}).}
\end{align}
Since we store all of the edges in $R_1$ in our memory, we can find a matching  of size $\mu(R_1)$ in $G_P$ such that the edges of this matching do not interfere with the edges in $\match^*_2$. Each matching edge in this subgraph will be a lower wing for an edge in $\match_P$. Therefore, the algorithm finds $\mu(R_1)$ augmenting paths. These augmenting paths interfere with at most $|R_P|$ edges of $\match^*_2$. Using these augmenting paths, we can find a matching of size at least
\begin{align}
\label{ieq:augp}
|\match_0|+ \mu(R_1) - |R_P| + |\match^*_2| \,.
\end{align}
Similarly, using the edges in $R_2$, we can get a matching of size at least
\begin{align}
\label{ieq:augq}
|\match_0|+ \mu(R_2) - |R_Q| + |\match^*_2| \,.
\end{align}
We show that the summation of sizes of these matchings is at least $\mu(G) \big(6/5 - o(1)\big)$, therefore the size of at least one of them is $\mu(G) \big(3/5 - o(1)\big)$. By taking the summation of (\ref{ieq:augp}) and (\ref{ieq:augq}) we get
\begin{align*}
&2|\match_0|+ \mu(R_1) + \mu(R_2)- |R_P| -|R_Q| + 2|\match^*_2| \nonumber\\
&\ge 2|\match_0|+ 2|\match^*_2| - \mu(G) \big(1/5+o(1)\big) \nonumber \\
&\quad+|P_1| + |Q_1| - |R_P| -|R_Q|  &\text{By (\ref{ieq:mur1r2}).} \nonumber\\
&\ge 2|\match_0| + 2|\match^*_2| + \mu(G) \big( \alpha - 2\alpha \delta - 4/5 - o(1)\big) \nonumber &\text{By the assumption of Case 3.} \\
&= (1+ 2 \delta)|\match^*_1| + 2|\match^*_2| + \mu(G) \big( \alpha - 2\alpha \delta - 4/5 - o(1)\big) \nonumber &\text{Since $|\match_0|=(1/2+\delta) |\match^*_1|$.} \\
&= 2|\match^*_2| + \mu(G) \big( 2\alpha - 4/5 - o(1)\big) \nonumber &\text{Since $|\match^*_1|=\alpha |\match^*|$.} \\
&= \mu(G) \big( 2 - 4/5 - o(1)\big) \nonumber &\text{Since $|\match^*_2|=(1-\alpha) |\match^*|$.} \\
&=\mu(G) \big(6/5 - o(1)\big) \,.
\end{align*}
Therefore, the size of the matching found by the algorithm is at least $\mu(G) \big(3/5 -o(1)\big)$ which is the desired bound. This completes the analysis for the last remaining case and completes the proof of the theorem.
\end{itemize}
\end{proof}

\section{Reduction from General Graphs to Bipartite Graphs}

\begin{algorithm} [h]
 \KwData{A random order stream $S=\langle e_1, \ldots, e_m \rangle$ of edges of a bipartite graph $G=(A,B,E)$, and a semi-streaming algorithm $\alg$ for finding a matching in bipartite graphs.}
 \begin{algorithmic} [1]
 \STATE Run $\gr$ on the edges $S_{[1,m/\log n]}$ to find a matching $\match_0$.
 \STATE Run the following subroutines in parallel.
 \STATE \hspace{\algorithmicindent} \textbf{Subroutine 1:}
  \STATE \hspace{\algorithmicindent}\hspace{\algorithmicindent} Let $G'=(V(\match_0), \overline{V(\match_0)})$ be a bipartite graph such that an each $e$ is in $G'$ if exactly one of its endpoints is in $V(\match_0)$.
 \STATE \hspace{\algorithmicindent}\hspace{\algorithmicindent} Run $\alg$ to find a matching $\match_1$ using the stream of edges $e \in S_{(m/\log n,m]}$ such that $e \in E(G').$
 \STATE \hspace{\algorithmicindent} \textbf{Subroutine 2:}

 \STATE \hspace{\algorithmicindent}\hspace{\algorithmicindent} Store all edges  $e \in S_{(m /\log n , m]}$ such that $e \cap V(\match_0) = \O$. Let $R$ be the set of these edges.
 \RETURN $\ms(\match_0 \cup R \cup \match_1)$.
 \end{algorithmic}
\caption{Algorithm $\gm$ for finding an approximate matching in general graphs.}
\label{alg:maingen}

\end{algorithm}

In this section, we present a black-box reduction from the finding a maximum matching in general graphs to that of the finding a maximum matching in bipartite graphs when edges arrive in a uniformly random order. The algorithm is formalized in Algorithm \ref{alg:maingen}. Algorithm \ref{alg:maingen} gets a graph $G=(V,E)$, and an arbitrary semi-streaming algorithm for finding a maximum matching in bipartite graphs. Assuming that the approximation ratio of the bipartite algorithm is $p$ when edges arrives in a random order, Algorithm \ref{alg:maingen} is guaranteed to have an approximation ratio of $\frac{2p}{2p+1}$ w.h.p. in $n$ where $n$ is the number of vertices in $G$. Using the fact that there exists a single-pass semi-streaming algorithm that finds $3/5-o(1)$ approximation of maximum matching in bipartite graphs w.h.p. in $n$, this immediately implies the following theorem.

\mainthree*

Now we show that the approximation ratio of our reduction method is $\frac{2p}{2p+1}$.

\maintwo*
\begin{proof}
The algorithm picks the first $1/\log n$ fraction of the edges and runs the greedy algorithm to find a matching $\match_0$. Let $R$ be the set of edges in the graph that do not interfere with the edges in $\match_0$, i.e., $R=E(G \setminus V(\match_0) )$. By Lemma \ref{lem:smallremaining}, we have that w.h.p., the number of edges in $R$ is bounded by $\tilde{O}(n)$ and all of them can be stored within the memory bound of semi-streaming algorithms. Consider an arbitrary maximum matching $\match^*$ among the remaining $(1-1/\log n)$ fraction of the edges. Each edge $e \in \ms(G)$ is in the last $(1-1/\log n)$ fraction of the edges with the probability of $(1-1/\log n)$. Therefore, $\E\big[|\match^*|\big] \ge (1-o(1)) \ms(G)$. Recall that we can assume $\ms(G) \ge \sqrt{n}$. In that case, with the standard application of Chernoff bound it can be shown that w.h.p. in $n$, we have $|\match^*| \ge \big(1-o(1)\big) \ms(G)$. Let $\match^*_1$ be the set of edges in $\match^*$ that are incident to the edges in $\match_0$ and $\match^*_2$ be the set of all other edges. Let $|\match^*_1| = \alpha \ms(G)$. It implies that
\begin{align}
\label{ieq:match2}
|\match^*_2| = |\match^*| - |\match^*_1| \ge (1-\alpha - o(1)) \ms(G) \,.
\end{align}
Since $\match_0$ is a maximal matching in the graph $\match^*_1$, we have $|\match_0| = (1/2+ \delta) |\match^*_1|$ where $0 \le \delta \le  1/2$. Since our semi-streaming algorithm can store all of the edges in $R$, it can output $\ms(\match_0 \cup R)$ as a matching. The size of this matching is at least,
\begin{align*}
\ms(\match_0 \cup R) &=|\match_0| + \ms(R) \\
&\ge |\match_0| + |\match^*_2| \\
& \ge  |\match_0| +  (1-\alpha - o(1)) \ms(G) & \text{By (\ref{ieq:match2}).}\\
& = (1/2+ \delta) |\match^*_1| +  (1-\alpha - o(1)) \ms(G) \\
& =  (1/2+ \delta) \alpha \ms(G) +  (1-\alpha - o(1)) \ms(G) & \text{Since $|\match^*_1| = \alpha \ms(G)$.} \\
& = \ms(G) (1-\alpha/2 +\alpha \delta -o(1)) \,.
\end{align*}
Consider the case that $\frac{\alpha}{2} - \alpha \delta \le \frac{1}{1+2p}$, in that case we have
\begin{align*}
\ms(\match_0 \cup R) & \ge \ms(G) \big(1-\alpha/2 +\alpha \delta -o(1)\big) \\
& \ge \ms(G) \big(1-\frac{1}{1+2p} -o(1)\big)\\
& = \ms(G) \big(\frac{2p}{1+2p} -o(1)\big)\,.
\end{align*}
Which is our desired bound. 

The remaining case is when $\frac{\alpha}{2} - \alpha \delta > \frac{1}{1+2p}$. In this case, the algorithm creates a bipartite graph between the matched vertices and unmatched vertices by $\match_0$. Specifically, let $G'=(V(\match_0), \overline{V(\match_0)})$ be a bipartite graph that has the matched vertices on one of its side and unmatched vertices on the other side. We add an edge $e$ to $G'$ if it is between matched and unmatched vertices. We claim that $\ms(G') \ge 2|\match^*_1| - 2|\match_0|$. To show that our claim holds, note that $\match_0$ is a maximal matching among the edges in $\match^*_1$. There are $2|\match_0|$ vertices that are covered by $\match_0$ and $2|\match^*_1|$ vertices that are covered by $\match^*_1$. Consider $2|\match^*_1| - 2|\match_0|$ vertices that are covered by $\match^*_1$ but are not covered by $\match_0$. Since $\match_0$ is maximal, each edge in $\match^*_1$ that covers these vertices is between $V(\match_0)$ and $\overline{V(\match_0)}$. Therefore, $\ms(G') \ge 2|\match^*_1| - 2|\match_0|$. Let $\alg$ be the algorithm for the bipartite matching. Since the approximation ratio of $\alg$ is $p$, using this algorithm we can find $p$ approximation of the maximum matching in $G'$. Let $\match_1$ be the matching returned by $\alg$ in $G'$. Then, we have
\begin{align*}
|\match_1| &\ge p \cdot \ms(G') \\
&\ge p\big(2|\match^*_1| - 2|\match_0|\big) \\
& = 2p\big(|\match^*_1| - |\match_0|\big) \\
& = 2p\big(\alpha \ms(G) - |\match_0|\big) & \text{Since $|\match^*_1|= \alpha \ms(G)$.}\\
& = 2p\big(\alpha \ms(G) - (1/2+\delta) \alpha \ms(G) \big) & \text{Since $|\match_0| = (1/2+\delta) |\match^*_1|$.} \\
& = 2p \cdot \ms(G) \big(\alpha/2 - \alpha \delta \big) \\
& > 2p \cdot \ms(G) \big(\frac{1}{1+2p} \big) & \text{Since $\alpha/2 - \alpha \delta > \frac{1}{1+2p}$.} \\
& = \frac{2p}{1+2p} \ms(G) \,,
\end{align*}
which is the desired bound and completes the proof for the reduction.
\end{proof}

\bibliographystyle{apalike} 
\bibliography{matching}
\newpage

\appendix
\section{Omitted proofs}\label{sec:appx}

\matchsides*
\begin{proof}
Pick any arbitrary maximum matching $\match^*$ in $G$. This matching covers exactly $\ms(G)$ vertices of $A$ and exactly $\ms(G)$ vertices of $B$. Similarly, the matching $\match$ covers exactly $|\match|$ vertices in both $A$ and $B$. Therefore, there are at least $\ms(G) - |\match|$ vertices in $A$ that are covered by $\match^*$ but are not covered by $\match$. Since $\match$ is maximal, the edges in $\match^*$ that cover these vertices are incident to the edges in $\match$. Therefore, these edges are between $B(\match)$ and $\overline{A(\match)}$. Thus, the size of a maximum matching between $B(\match)$ and $\overline{A(\match)}$ is at least
\begin{align*}
\ms(G) - |\match| = (1/2- \delta) \ms(G) \,.
\end{align*}
Using a similar argument it can also shown that the size of maximum matching between $A(\match)$ and $\overline{B(\match)}$ is also at least $(1/2- \delta) \ms(G)$.
\end{proof}

\matchminus*
\begin{proof}
We know that there exists a matching of size $(1/2- \delta) |\match^*_1|$ in $G_L$. Therefore the number of vertices in $B(\match_P)$ that are not matched to a vertex in $A(G_L)$ is bounded by
\begin{align*}
|\match_0| - (1/2- \delta) |\match^*_1|&= (1/2+\delta) |\match^*_1| - (1/2- \delta) |\match^*_1|\\
&=2\delta |\match^*_1| = \ms(G) \cdot 2\delta \alpha \,.
\end{align*} 
Therefore, there are at most $\ms(G) \cdot 2\delta \alpha$ vertices in $B(\match_P)$ that cannot be matched which proves the claim.
\end{proof}

\end{document}